%% file: main.tex
\newif\ifreview
\reviewfalse
   
\ifreview
  \documentclass[sigconf, screen,review=false,nonacm, acmthm=false,natbib=false]{acmart}
\else
  \documentclass[sigconf,natbib=false,screen,acmthm=false]{acmart}
\fi

\usepackage[T1]{fontenc}
\usepackage[utf8]{inputenc}
\usepackage[english]{babel}
\usepackage[autostyle]{csquotes}

\RequirePackage[abbreviate=true, dateabbrev=true, isbn=true, doi=true, urldate=comp, url=true, maxbibnames=9, backref=false, backend=biber, style=ACM-Reference-Format, language=american, sorting=nyvt]{biblatex}

\newcommand*{\origrightarrow}{}

\renewcommand*{\textrightarrow}{\fontfamily{cmr}\selectfont\origrightarrow}

\usepackage{booktabs} 

\setcopyright{none}

\usepackage{mathptmx} 

\usepackage{soul}
\usepackage{amsthm}
\usepackage{amsmath}
\usepackage{latexsym}
\usepackage{amssymb}
\usepackage{cmll}
\usepackage{subfig}
\usepackage{algorithmicx}
\usepackage{array}
\usepackage{listings}
\usepackage{mathtools}
\usepackage[english]{babel}
\usepackage{blindtext}
\usepackage{fancyhdr}
\usepackage{hyperref}
\usepackage{xspace}
\usepackage{algorithm}
\usepackage{mathptmx}
\usepackage[noend]{algpseudocode} 
\usepackage{wrapfig}
\usepackage{listings}
\usepackage[fulladjust]{marginnote}
\usepackage[normalem]{ulem}
\usepackage{flushend}
\usepackage{color}
\usepackage{xargs}                      
\usepackage[normalem]{ulem}

\DeclareRobustCommand*\circled[1]{\tikz[baseline=(char.base)]{
            \node[shape=circle,white,draw,fill=black,inner sep=1.4pt] (char) {#1};}}

\newcommand{\ignore}[1]{}

\algnewcommand{\IIf}[1]{\State\algorithmicif\ #1\ \algorithmicthen}
\algnewcommand{\EndIIf}{\unskip\ \algorithmicend\ \algorithmicif}

\DeclarePairedDelimiter\floor{\lfloor}{\rfloor}

\usepackage[colorinlistoftodos,prependcaption,draft,textsize=tiny]{todonotes}

\newif\ifsubmit
\submitfalse
\ifsubmit
    \newcommand{\usure}[1]{}
    \newcommand{\change}[1]{}
    \newcommand{\info}[1]{}
    \newcommand{\improvement}[1]{}
    \newcommandx{\cheng}[2][1=]{}
    \newcommandx{\abdul}[2][1=]{}
    \newcommandx{\simon}[2][1=]{}
    \newcommandx{\junjun}[2][1=]{}
    \newcommandx{\wenmei}[2][1=]{}
\else
	 
	\newcounter{adtodocounter} 
	\newcounter{cltodocounter} 
	\newcounter{jjtodocounter} 
	\newcounter{wmhtodocounter} 
	\newcounter{sgtodocounter} 
    \newcommandx{\unsure}[2][1=]{\todo[linecolor=red,backgroundcolor=red!25,bordercolor=red,#1]{#2}}
    \newcommandx{\change}[2][1=]{\todo[linecolor=blue,backgroundcolor=blue!25,bordercolor=blue,#1]{#2}}
    \newcommandx{\info}[2][1=]{\todo[linecolor=OliveGreen,backgroundcolor=OliveGreen!25,bordercolor=OliveGreen,#1]{#2}}
    \newcommandx{\improvement}[2][1=]{ \marginpar[\todo[linecolor=Plum,backgroundcolor=Plum!25,bordercolor=Plum,#1]{#2}]{}}
    \newcommandx{\thiswillnotshow}[2][1=]{\todo[disable,#1]{#2}}
    \newcommandx{\cheng}[2][1=]{\stepcounter{cltodocounter} \todo[linecolor=red,backgroundcolor=purple!25,bordercolor=purple,#1]{CL(\thecltodocounter): #2}}
    \newcommandx{\abdul}[2][1=]{\stepcounter{adtodocounter} \todo[linecolor=red,backgroundcolor=red!25,bordercolor=red,#1]{AD(\theadtodocounter): #2}}
    \newcommandx{\simon}[2][1=]{\stepcounter{sgtodocounter} \todo[linecolor=OliveGreen,backgroundcolor=OliveGreen!25,bordercolor=OliveGreen,#1]{SG(\thesgtodocounter): #2}}
    \newcommandx{\jinjun}[2][1=]{\stepcounter{jjtodocounter} \todo[linecolor=yellow,backgroundcolor=yellow!25,bordercolor=yellow,#1]{JJ(\thejjtodocounter): #2}}
    \newcommandx{\wenmei}[2][1=]{\stepcounter{wmhtodocounter} \todo[linecolor=Plum,backgroundcolor=Plum!25,bordercolor=Plum,#1]{WMH(\thewmhtodocounter): #2}}
\fi

\definecolor{dkgreen}{rgb}{0,0.6,0}
\definecolor{gray}{rgb}{0.5,0.5,0.5}
\definecolor{mauve}{rgb}{0.58,0,0.82}

\makeatletter
\let\OldStatex\Statex
\renewcommand{\Statex}[1][3]{%
  \setlength\@tempdima{\algorithmicindent}%
  \OldStatex\hskip\dimexpr#1\@tempdima\relax}
\makeatother

\newcommand{\mpu}{TCU\xspace}
\newcommand{\mpus}{TCUs\xspace}

\usepackage{enumitem}
\setlist{nolistsep}

\copyrightyear{2019}
\acmYear{2019}
\setcopyright{acmlicensed}
\acmConference[ICS '19]{ACM/SPEC International Conference on Supercomputing}{June 26--28, 2019}{Pheonix, AX}
\acmBooktitle{ACM/SPEC International Conference on Supercomputing (ICS '19), June 26--28, 2019, Pheonix, AX}
\acmPrice{15.00}
\acmDOI{10.1145/3330345.3331057}
\acmISBN{978-1-4503-6079-1}

\begin{document}

\settopmatter{printfolios=true,printacmref=true}
\fancyhead{}

\title{Accelerating Reduction and Scan Using Tensor Core Units}

\ifreview
\else

\author{Abdul Dakkak, Cheng Li}
\affiliation{%
  \institution{University of Illinois Urbana-Champaign}
  \city{Urbana}
  \state{Illinois}
  \postcode{61801}
}
\email{{dakkak, cli99}@illinois.edu}

\author{Jinjun Xiong}
\affiliation{%
  \institution{IBM T. J. Watson Research Center}
  \city{Yorktown Heights}
  \state{New York}
}
\email{jinjun@us.ibm.com}

\author{Isaac Gelado}
\affiliation{%
  \institution{NVIDIA Corporation}
  \city{Santa Clara}
  \state{California}
}
\email{igelado@nvidia.com}

\author{Wen-mei Hwu}
\affiliation{%
  \institution{University of Illinois Urbana-Champaign}
  \city{Urbana}
  \state{Illinois}
}
\email{w-hwu@illinois.edu}

\renewcommand{\shortauthors}{A. Dakkak et al.}

\fi

\input{sec/0-abstract.tex}




\maketitle

\input{sec/1-intro.tex}
\input{sec/2-background.tex}
\input{sec/3-impl.tex}
\input{sec/4-reduction.tex}
\input{sec/5-prefixsum.tex}
\input{sec/6-evaluation.tex}

\input{sec/8-related.tex}
\input{sec/9-conclusion.tex}

\input{sec/99-ack.tex}

\printbibliography

\end{document}

%% file: sec/0-abstract.tex
\begin{abstract}

Driven by deep learning, there has been a surge of specialized processors for matrix multiplication, referred to as Tensor Core Units~(\mpus).
These \mpus are capable of performing matrix multiplications on small matrices (usually $4 \times 4$ or $16 \times 16$) to accelerate HPC and deep learning workloads.
Although \mpus are prevalent and promise increase in performance and/or energy efficiency, they suffer from over specialization as only matrix multiplication on small matrices is supported.
In this paper we express both reduction and scan in terms of matrix multiplication operations and map them onto \mpus.
To our knowledge, this paper is the first to try to broaden the class of algorithms expressible as \mpu operations and is the first to show benefits of this mapping in terms of: program simplicity, efficiency, and performance.
We implemented the reduction and scan algorithms using NVIDIA's V100 \mpus and achieved $89\%-98\%$ of peak memory copy bandwidth.
Our results are orders of magnitude faster (up to $100\times$ for reduction and $3\times$ for scan) than state-of-the-art methods for small segment sizes (common in HPC and deep learning applications).
Our implementation achieves this speedup while decreasing the power consumption by up to $22\%$ for reduction and $16\%$ for scan.
\end{abstract}

%% file: sec/1-intro.tex
\section{Introduction}\label{sec:introduction}

Deep learning's reliance on matrix-multiplication (GEMM) for compute has driven both research and industry to develop matrix-multiplication accelerator hardware --- collectively called Tensor Core Units~(\mpus) in this paper.
\mpus are designed to accelerate Multilayer Perceptrons~(MLP), Convolutional Neural Networks~(CNN), and Recurrent Neural Networks~(RNN)
or Deep Neural Network (DNN) in general.
\mpus come under the guise of different marketing terms, be it NVIDIA's Tensor Cores~\cite{tensorcores}, Google's Tensor Processing Unit~\cite{cloudtpu}, Intel's DLBoost~\cite{cascadelake}, Apple A11's Neural Engine~\cite{apple}, Tesla's HW3, or ARM's ML Processor~\cite{arm}.
They vary in the underlying hardware implementation~\cite{du2017accelerator,jouppi2017datacenter,reagen2017deep,zhu2018mobile}, and are prevalent~\cite{edge-tpu,tensorcores,parssinen2018modern} in both cloud and edge devices.

\begin{figure}[t]
	\centering
	\includegraphics[width=0.44\textwidth]{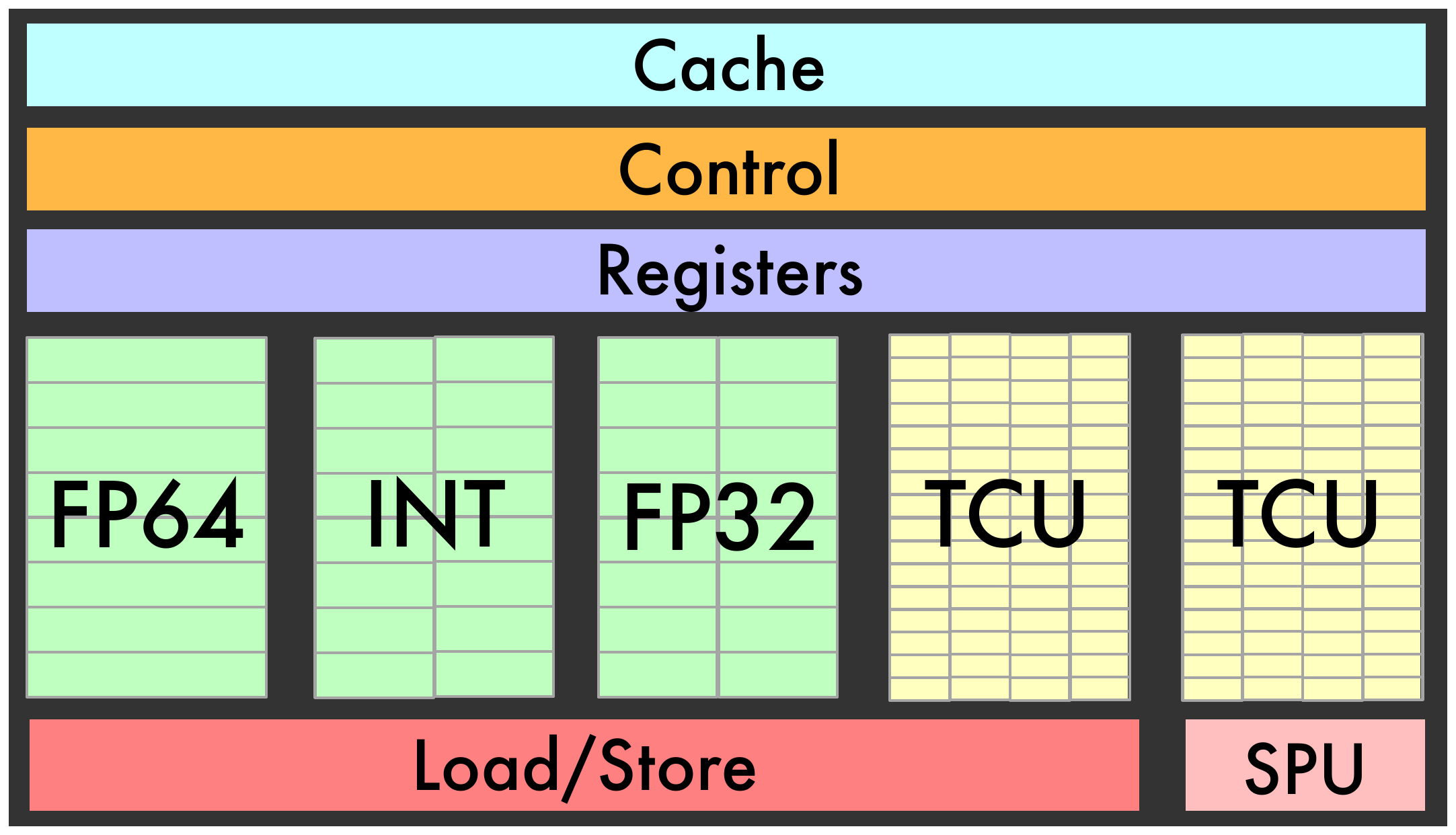}
	\caption{Each subcore (processing block) in the NVIDIA Tesla V100 PCI-E architecture contains $2$ \mpus. In total, $640$ \mpus are available --- achieving a theoretical peek of $113$ TFLOPS.
	}
	\label{fig:v100}
\end{figure}

To show the theoretical benefits of \mpus, consider the NVIDIA Volta V100 GPUs architecture. 
Using V100 Tensor Cores, one achieves a $8\times$ throughput increase per Streaming Multiprocessors~(SM) over previous Pascal GP100 generation.
This throughput increase is because each  V100 SM  is capable of performing $1024$ half precision operations per cycle using the \mpus whereas the GP100 SM is  capable of performing $128$ half precision operations per cycle without the \mpus.
The throughput increase is enabled by the fact that the V100  dedicates a large chip area of the SM subcore  to \mpus (Figure~\ref{fig:v100}).

Although \mpus are prevalent and promise increase in performance and/or energy efficiency and are heavily used within supercomputers~\cite{sierra,summit} to achieve exascale performance, they suffer from over specialization.
Currently, no algorithm other than GEMM utilizes the NVIDIA \mpus.
This results in idle \mpus, low chip utilization, and limits \mpus applicability to  specialized libraries or narrow application domains.

The objective of the paper is to expand the class of algorithms that can execute on \mpus --- enabling the \mpus to be used within a wider range of non-GEMM algorithms. 
We choose reduction and scan, since a large body of work~\cite{blelloch1993segmented,chan2010more,mccool2012structured} has shown that they are key primitives for data parallel implementations of  radix sort, quicksort, lexical analysis, stream compaction, and polynomial evaluation.
In this paper, we formulate a mapping of reduction or scan onto \mpus.
We then introduce algorithms for cache- (warp-), processing element (PE)/core- (block-), and device- (grid-) level reduction and scan and show their performance on NVIDIA \mpus.
We separate our algorithm description from implementation, making the algorithms, motivation, methods, and observations generally applicable to a broader range of \mpus and numerical precision agnostic.
While the formulation is the main objective of the paper, we show that an implementation of our algorithms on NVIDIA V100 is either order of magnitude faster  or rival the fastest GPU implementation, with much lower programming complexity.
The key contributions of the paper are:

\begin{enumerate} 
	\item We show how to use \mpus to compute both reduction and scan. We believe we are the first to formulate these algorithms in terms of \mpu operations in a manner that is independent to the underlying \mpu architecture.
	\item We implement our algorithms onto NVIDIA V100 GPUs and show orders of magnitude speedup over state-of-art algorithms for small segment sizes. Small segements are common in mathematics (e.g. evaluating polynomials), scientific applications (e.g. finite difference), and machine learning (e.g. batch norm) applications. For large segments, we are comparable to the fastest algorithms and achieve $89-98\%$ of theoretical peak memory copy bandwidth.
	\item We show that our implementation is up to $22\%$ more power efficient and decreases the utilization of general purpose ALUs.
	\item We describe the current usage and programmability of the NVIDIA TensorCore and evaluate GEMM on the \mpus using cuBLAS~\cite{cublas}, CUTLASS~\cite{cutlass} and the CUDA \mpu API.
\end{enumerate}

This paper is divided as follows: we first describe the NVIDIA \mpus and show the performance of GEMM and GEMV computation in Section~\ref{sec:background}. 
In Section~\ref{sec:current-impl}, we give a background of reduction and scan and show the \mpu algorithms for reduction (Section~\ref{sec:reduction}) and scan (Section~\ref{sec:prefixsum}).
We then compare our implementation against state-of-the-art in Section~\ref{sec:evaluation}. 
Section~\ref{sec:related} describes the related work, before we conclude in Section~\ref{sec:conclusion}.

%% file: sec/2-background.tex

\section{Tensor Cores Units (TCUs)}\label{sec:background}

A marquee feature of NVIDIA's GPUs (Volta's Tesla V100 and Turning's TU102 architectures) and Google's TPUs are their \mpus --- a programmable matrix multiply and accumulate hardware units, called Tensor Cores
by NVIDIA and matrix-multiply-units~(MXUs) by Google \footnote{We will use \mpu and Tensor Core interchangeably in this paper.}. 
While there are other competing \mpu implementations, both NVIDIA Tensor Cores and Google's TPU are by far the most popular.
At a high level, their functionality and architectural design are similar.
They both subdivide the device into cores, with each having multiple processing block (or subcores) and \mpus.
Figure~\ref{fig:v100} illustrates a subcore in an NVIDIA SM, with the V100 containing $80$ SMs and each having $4$ subcores.
In turn, each subcore contains two Tensor Cores --- for a total of $640$ Tensor Cores and achieve a $12 \times$ throughput improvement over previous generation Tesla P100~\cite{tensorcorescuda9}.
Google's TPUv3 device, on the other hand, has $8$ cores --- $4$ chips each with $2$ cores --- with each core having $2$ MXUs.

Since Google TPUs currently can only be used within Google Cloud using the XLA compiler~\cite{tfxla} and the NVIDIA V100 \mpus are widely available and are installed in supercomputers~\cite{sierra,summit}, this section will only describe the \mpu usage and results for NVIDIA V100.
Similar analysis can be performed for other \mpus.

\begin{figure*}[ht!]   
\centering
\subfloat[GEMM with half precision input and half precision output.]{
\includegraphics[width=0.5\textwidth, keepaspectratio]{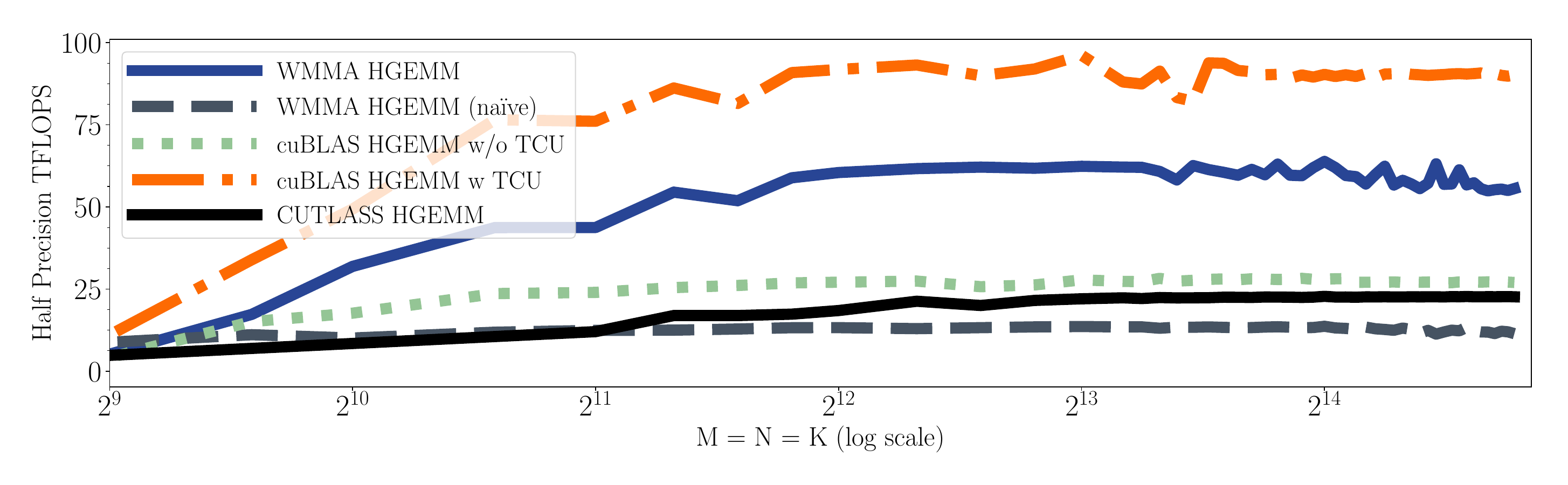}
\label{fig:gemm_16_16}}
\subfloat[Mixed precision GEMM with half precision input and single precision output.]{
\includegraphics[width=0.5\textwidth, keepaspectratio]{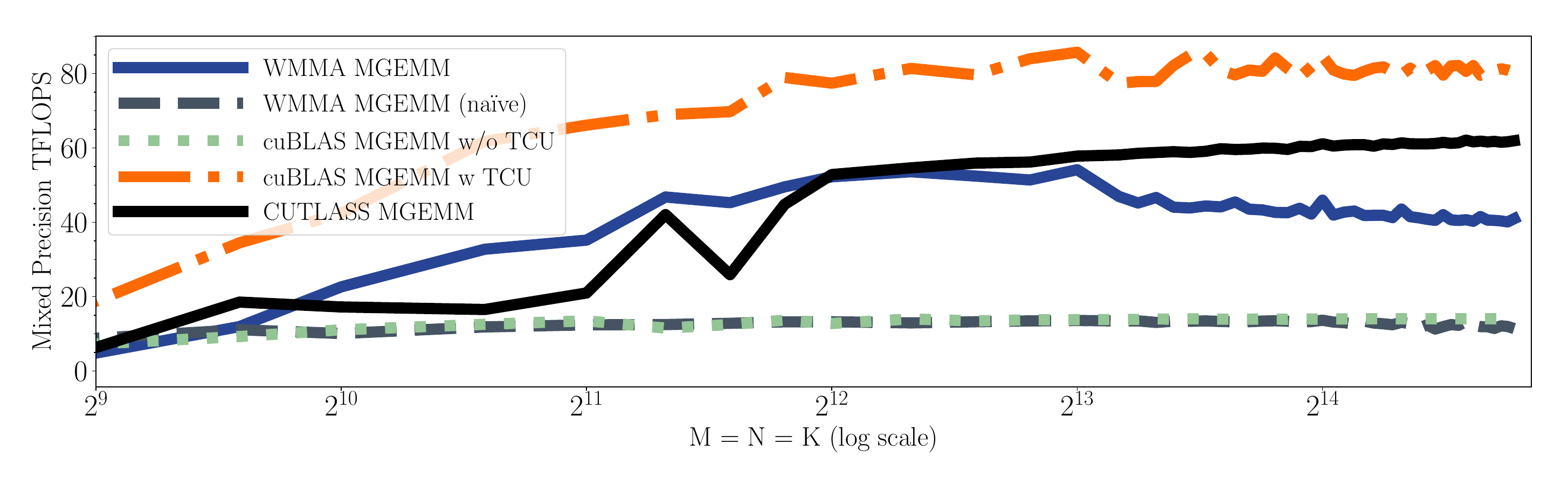}\label{fig:gemm_16_32}}
\caption{General matrix-matrix multiplication (GEMM) performance using Tensor Cores for both half-~(\ref{fig:gemm_16_16})  and mixed-~(\ref{fig:gemm_16_32}) precision on a V100 PCI-E GPU with a clock frequency of $1380$ MHz and a $113$ TFLOPS peek performance.
The inputs are square matrices with variable $\langle M , N , K \rangle$ dimensions. 
The optimized and na\"ive WMMA GEMM algorithms are described in the text.
}
\label{fig:gemm}
\end{figure*}


Each NVIDIA V100 Tensor Core provides a $4 \times 4 \times 4$ tensor processing array capable of performing the operation $D = A \cdot B + C$ within a cycle, where $A$, $B$, $C$ and $D$ are $4 \times 4$ matrices.
The inputs $A$ and $B$ must be in half precision format while the accumulators, $C$ and $D$, can be either single or half precision.
Each Tensor Core can perform  $64~\left (4\times4\times4 \right )$ FMA operations per cycle.
Therefore, using the \mpu each SM can perform $1024~\left (64\times2\times8 \right )$ floating point operations per cycle, since each FMA consists of two floating point operations and each SM contains $8$ Tensor Cores.
This is an $8\times$ SM throughput increase compared to Pascal for floating point operations~\cite{tensorcorescuda9}.
This section first describes the current usage of the NVIDIA Tensor Cores, then details the current NVIDIA Tensor Cores API, and presents evaluation results to motivate our work.

\subsection{Current Library Usage}
Currently, Tensor Cores have only been used to accelerate GEMM operations, most prominently through NVIDIA's CUDA libraries (such as cuBLAS~\cite{cublas} and cuDNN~\cite{cudnn}).
These libraries require users to opt-in to use the Tensor Cores to accelerate GEMM computation.
NVIDIA also provides the CUTLASS (CUDA Templates for Linear Algebra Subroutines)~\cite{cutlass} library, which is a C++  templated library that provides building block primitives to write high performance GEMM-like kernels.
Deep learning frameworks such as NVCaffe~\cite{jia2014caffe}, Caffe2~\cite{caffe2}, MXNet~\cite{mxnet}, PyTorch~\cite{pytorch},  TensorFlow~\cite{tensorflow}, and  TensorRT~\cite{tensorrt} leverage these NVIDIA libraries for DNN training~\cite{mixedprecisiontraining} and inference acceleration.

\subsection{Programming Interface}
Aside from the libraries, NVIDIA also provides a CUDA C++ Warp Matrix Multiply and Accumulate (WMMA)~\cite{wmma} API to program the Tensor Cores directly.
The current WMMA API provides warp-level matrix operations for matrix load (\texttt{load\_matrix\_sync}), matrix store (\texttt{store\_matrix\_sync}), and matrix multiply and accumulate (\texttt{mma\_sync}).
These APIs operate on a special
data type \texttt{fragment}, which holds a matrix tile in thread-local registers.
A helper function to broadcast a scalar constant into a fragment (\texttt{fill\_fragment}) is provided as well.
No API currently exists for calling \mpu operations at sub warp level --- neither in the IR nor in the PTX~\cite{nvvmir,ptx}.

The \texttt{load\_matrix\_sync} function distributes values of the matrix across the warp lanes.
Threads within a warp utilize multiple Tensor Cores concurrently to perform the \texttt{mma\_sync} operation --- collaborating to compute the  $D_{M \times N} = A_{M\times K} \cdot B_{K \times N} + C_{M \times N}$, with $M$, $N$, $K$ denoting the matrix dimensions.
The API imposes limitations on the dimensions 
--- requiring the shape $\langle M,N,K \rangle$ to be either $\left \langle 16,16,16 \right \rangle$, $\langle 32,8,16 \rangle$, or $\left \langle 8,32,16 \right \rangle$.

\begin{lstlisting}[
  float=htp,
  floatplacement=htbp,
  basicstyle=\fontsize{6}{6}\ttfamily,
  caption={A simple CUDA kernel performing  $\left \langle 16, 16, 16 \right \rangle$ matrix multiplication ($C = A \cdot B + C$) in half precision using the CUDA WMMA API.},
  frame=lines,
  numbers=left,
  xleftmargin=2em,
  framexleftmargin=3em,
  stepnumber=1,
  escapechar=|,
  label=lst:wmma_dot,
  captionpos=b]
#include <mma.h>
using namespace nvcuda::wmma;
__global__ void dot_wmma_16x16(half *a, half *b, half *c) {
   |\label{line:start_def_frag}|fragment<matrix_a, 16, 16, 16, half, col_major> a_frag; 
   fragment<matrix_b, 16, 16, 16, half, row_major> b_frag;
   fragment<accumulator, 16, 16, 16, half> c_frag;  |\label{line:end_def_frag}|
   load_matrix_sync(a_frag, a, /* row stride */ 16);  |\label{line:start_load_frag}|
   load_matrix_sync(b_frag, b, /* row stride */ 16);  |\label{line:end_load_frag}|
   fill_fragment(c_frag, 0.0f);  |\label{line:fill_frag}|
   mma_sync(c_frag, a_frag, b_frag, c_frag);  |\label{line:dot}|
   store_matrix_sync(c, c_frag, 16, row_major);  |\label{line:store_frag}|
}
\end{lstlisting}

Listing~\ref{lst:wmma_dot} shows a CUDA kernel that computes a $\left \langle 16,16,16 \right \rangle$ matrix multiplication within a warp using the WMMA API.
Lines~\ref{line:start_def_frag}--\ref{line:end_def_frag} declare the matrix fragments. 
The API supports $3$ kinds of matrices --- \texttt{matrix\_a} ($A$), \texttt{matrix\_b} ($B$), and \texttt{accumulator} ($C$ or $D$) --- with each having their own internal data layout~\footnote{The mapping between individual matrix elements to their residing thread(s) is purposely opaque~\cite{wmma} and undocumented. We discuss how we alleviate some of the constraints in Section~\ref{sec:unsafe}.} as well as loading, storing, and computing semantics.
Users specify both the data type and the $\left \langle M,N,K \right \rangle$ shape of the fragments.
For both the $A$ and $B$ kinds, users specify whether the matrix is in column-~or row-major order. 
Users also specify the stride between rows and load the data from either shared or global memory (Lines~\ref{line:start_load_frag}--\ref{line:end_load_frag}).
Line~\ref{line:fill_frag} initializes the \texttt{matrix\_c} elements to zero by broadcasting the scalar value $0$ into the fragment.
Once the data is loaded, users perform the matrix multiplication operation (Line~\ref{line:dot}) and store the results (Line~\ref{line:store_frag}).



The kernel in Listing~\ref{lst:wmma_dot} can be generalized to implement GEMM for arbitrary matrix dimensions in a manner similar to tiling matrix multiplication. 
For example, a naive implementation (referred to as \texttt{WMMA HGEMM (na\"ive)}) assigns a strip of $16$ rows from matrix $A$ and a strip of $16$ columns from matrix $B$ columns to each warp to compute a $16\times16$ tile of the output $C$.
Each warp iterates through the $A$ rows and $B$ columns by loading $16\times16$ tiles of $A$ and $B$ from global memory into the fragments using \texttt{load\_matrix\_sync}, then performing \texttt{mma\_sync}, and repeats.
After all rows of $A$ and columns of $B$ have been consumed, the warp uses \texttt{store\_matrix\_sync} to store the accumulated $C$ values into   global memory.
An optimized implementation (referred to as \texttt{WMMA HGEMM}) utilizes persistent threads where each thread block collaboratively loads multiple tiles of matrix $A$ and $B$ into shared memory (to facilitate tile re-use).
The tiles are then loaded into fragments and the \texttt{mma\_sync} operation is performed.

\begin{figure}[t]
    \centering
    \includegraphics[width=0.5\textwidth]{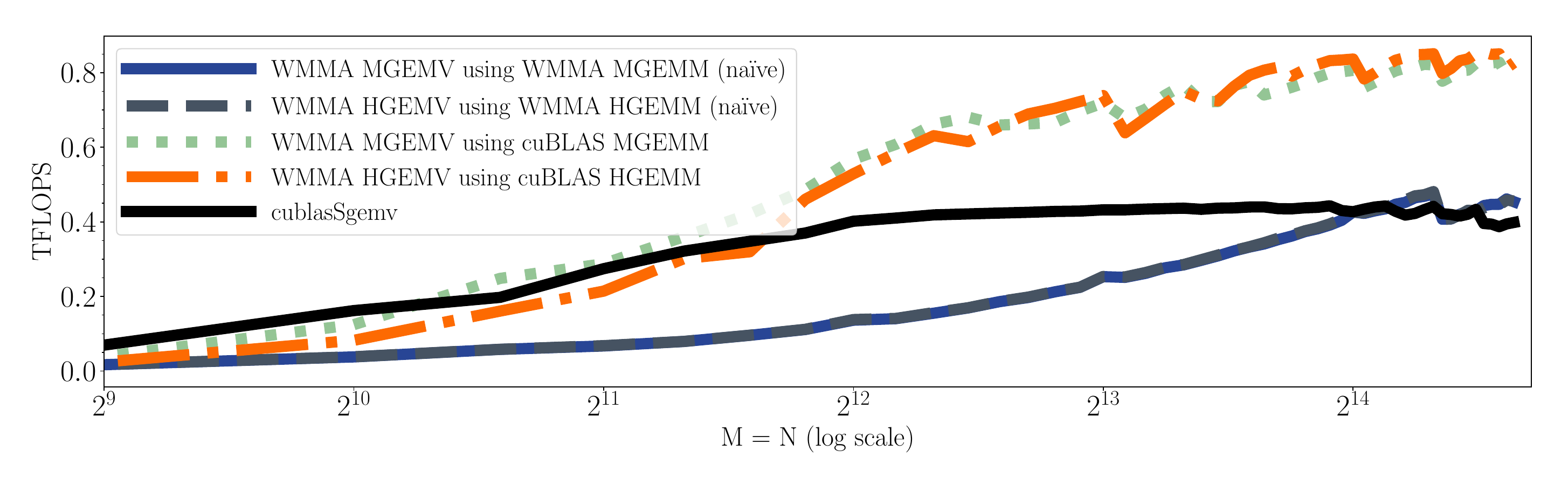}
    \caption{General matrix-vector multiplication (GEMV) performance using Tensor Cores on a V100 PCI-E GPU. GEMV can be implemented in terms of a GEMM (with dimensions $\langle M , N , 16 \rangle $) or calling the GEMV method in CUBLAS (which currently does not support half precision).}
    \label{fig:gemv}
\end{figure}

\subsection{GEMM Evaluation} 
To show the \mpu performance, we evaluate GEMM using Tensor Cores on an NVIDIA Tesla V100 PCI-E GPU with CUDA $9.2.88$ through cuBLAS, CUTLASS (version $0.1.1$), and hand written kernels using the WMMA API (Figure~\ref{fig:gemm}).

For half precision GEMM (HGEMM), shown in Figure~\ref{fig:gemm_16_16}, cuBLAS HGEMM with Tensor Cores achieves a maximum performance of $96.3$ TFLOPS --- approximately $85\%$ the peak performance --- and over $3.4 \times$ that of cuBLAS without the use of \mpus.
For mixed precision GEMM (MGEMM), shown in Figure~\ref{fig:gemm_16_32}, a maximum performance of $85.8$ TFLOPS is achieved on NVIDIA \mpus using cuBLAS, approximately $76\%$ the peak performance, for a $6.2\times$ speedup over cuBLAS without Tensor Cores (the degradation of performance compared to HGEMM is due to output bytes count being twice as large).
CUTLAS MGEMM is more performant than HGEMM, this is due to compiler and hardware optimizations for mixed precision that are absent from half precision~\cite{Raihan}.

\subsection{GEMV Evaluation}
The order of magnitude speedup of GEMM with \mpu raises the question: can we formulate other algorithms in terms of matrix multiplication and also benefit from the \mpu?
The most obvious algorithm is  matrix-vector multiplication (GEMV).
We implement HGEMV (half precision GEMV) and MGEMV (mixed-precision GEMV) using cuBLAS HGEMM or MGEMM with dimension $\langle M , N , (K=16) \rangle $.
This method wastes at least $15 N$ memory loads and performs $15M N$ extra flops.
We evaluate our implementations against cuBLAS SGEMV, since half precision GEMV is not present within cuBLAS.

Figure~\ref{fig:gemv} shows that even when accounting for both resource and computation waste, HGEMV, implemented using cuBLAS HGEMM with Tensor Cores, outperforms cuBLAS SGEMV by at least $2\times$ and saturates at $900$ GFLOPS due to the HBM2  global memory bandwidth.
Na\"ive~\footnote{Note that one implicitly performs tiling when utilizing the WMMA API.} HGEMV and MGEMV are super imposed atop each other since the overhead of using mixed-precision is dwarfed by the inefficient memory access. 
Both na\"ive versions still outperform cuBLAS' SGEMV for large inputs.


The GEMV evaluation shows that the performance of matrix multiplication on NVIDIA \mpus is high enough to tolerate resource and computation waste in algorithms.
Driven by this observation, we examine how to formulate two widely used primitives --- reduction and scan --- to utilize \mpus. 

%% file: sec/3-impl.tex
\section{Reduction and Scan on GPUs}\label{sec:current-impl}

\begin{figure}[t]
  \centering
  \includegraphics[width=0.45\textwidth]{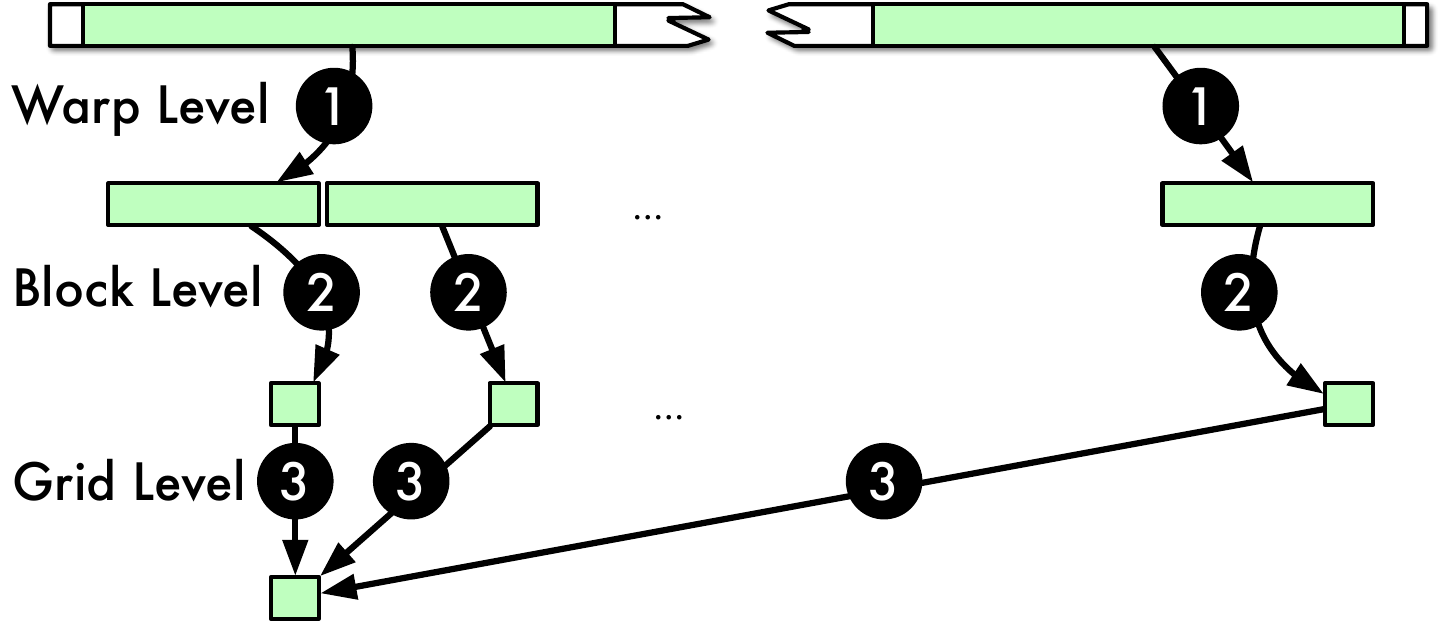}
  \caption{The reduction algorithm is \circled{1} composed of warp-level reduction that reduces each segment and is used to \circled{2} implement block-level reduction that further reduces each segment of partially reduced values. The partially reduced values are reduced across the grid \circled{3} to perform full reduction.}
  \label{fig:reduction_hierarchy}
\end{figure}




We start by defining reduction and scan.
Reduction (also called \textit{fold} or \textit{total}) of a vector  $A = \left[a_1, a_2, \ldots, a_n \right]$ is defined by its sum $\Sigma\sum_{i =1}^n a_i$.
Segmented reduction is defined as reductions on subsets of the input vector. 
In a regular segmented reduction, all segments are the same size\footnote{Irregular segmented reduction is implemented in terms of regular segmented reduction by padding the input.}. 
The \textit{scan} operation (also called \textit{prefix sum}) for the same vector $A$ is defined by the vector $\left[ a_1, a_1 + a_2, \ldots, \Sigma\sum_{i =1}^n a_i \right]$.
Segmented scan is defined similarly to segmented reduction.

\begin{lstlisting}[
  float=htp,
  floatplacement=tbp, 
  basicstyle=\fontsize{7}{7}\ttfamily,
  caption={
    NVIDIA's recommended warp-level reduction and scan implementations utilizing shuffle instructions.
  },
  label=lst:warp_collectives,
  frame=lines,
  numbers=left,
  xleftmargin=2em,
  framexleftmargin=3em,
  stepnumber=1,
  escapechar=|,
  captionpos=b]
__device__ half warp_reduce(half val) {
  for (int offset=WARP_SIZE/2; offset>0; offset/=2)
    val += __shfl_down_sync(0xFFFFFFFFU, val, mask);
  return val; }
__device__ half warp_scan(half val) {
  for (int offset=1; offset<WARP_SIZE; offset*=2) {
    auto n = __shfl_up_sync(0xFFFFFFFFU, val, mask);
    if (laneid >= offset) val += n; }
  return val; }
\end{lstlisting}

\subsection{State-of-the-art Implementations}

For GPUs, state of the art libraries~\cite{magma,openmp,cub} implement both reduction and scan in terms of warp-, block-, and device-level collectives, as illustrated in Figure~\ref{fig:reduction_hierarchy}.
The warp-level are commonly implemented using shuffle instructions~\cite{shuffle}, shown in Listing~\ref{lst:warp_collectives}, which allows threads within a warp to share values via registers without synchronization or using shared memory.
Shuffle instructions can be a bottleneck due to their limited throughput, however.
For example, on the NVIDIA Volta architecture only $32$ warp shuffle operations can be performed within a clock cycle per SM.

%% file: sec/4-reduction.tex
\section{\mpu Reduction Algorithm}\label{sec:reduction}

Intuitively, reduction can be implemented using \mpus by representing it as a special case of matrix multiplication, since 

\scalebox{.65}{\parbox{.5\linewidth}{%
\begin{align*}
\begin{aligned}
Red&uction(\left [ a_1, a_2, \ldots, a_n \right ] ) =
\begin{pmatrix}
1 & 1 & \cdots & 1 \\
0 & 0 & \cdots & 0 \\
\vdots & \vdots & \ddots & \vdots \\
0 & 0 & \cdots & 0
\end{pmatrix}\cdot%
\begin{pmatrix}
a_1 & a_2 & \ldots & a_n\\
0  & 0 & \cdots & 0\\
\vdots  & \vdots  & \ddots & \vdots\\
0  & 0 & \cdots & 0
\end{pmatrix}^T
= \begin{pmatrix}
\displaystyle\Sigma\sum_{i = 1}^n{ a_i} & 0 & \cdots & 0 \\
0 & 0 & \cdots & 0 \\
\vdots & \vdots & \ddots & \vdots \\
0 & 0 & \cdots & 0
\end{pmatrix}
\end{aligned}
\end{align*}
}}

The challenge is to map generic input sizes onto the fixed matrix dimensions supported by the \mpus.
For simplicity, this paper will assume that the \mpu supports only matrices with $16 \times 16$ dimension.
Other hardware may require other dimensions and those can be used without modifying the core idea of the algorithms.
The algorithms are also presented in a precision agnostic way.


We use $Reduction_K$ to represent a $K$ regular segmented reduction --- partial reductions of the input uniformly partitioned  into $K$ element subsets.
We will use $P$ to denote the matrix which has ones for the first row and zero otherwise (i.e. $ p_{r,c} = \begin{cases} 1 \quad \text{if } r = 0 \\ 0 \quad \text{if } r \neq 0 \end{cases}$), and 
 the notation  $\underline{\mathbf{X}}$ for a matrix where all elements are the constant value $X$.

To make our formulation non-NVIDIA WMMA API specific, we present our algorithms in an API neutral way.
In the following sections, we use \textbf{LoadTile} in place of the \texttt{load\_matrix\_sync} which takes a memory address, a matrix layout (default is row-major), and stride (default is $16$) as input.
We abstract \texttt{store\_matrix\_sync} to make it addressable as if it were a linear array.
We will also use the notation $A \cdot B + C$ to denote the \texttt{mma\_sync} operation. 
This paper however uses the standard CUDA terminology for warp, block, and grid to explain the algorithms, since no other standard nomenclature exists.
The warp, block and device used in this paper correspond to the three memory hierarchy levels: L-Cache, PE/core, and device.

\subsection{L-Cache (Warp)-level Reduction}

We introduce warp-level reduction first, since it is the building block for both block- and grid-level reductions.
We formulate reduction using \mpus for segment sizes $16$, $256$, and multiples of $16$ and $256$.
Support for arbitrary segment sizes can be supported either by padding the input with zeros or by masking the $P$ matrix.
We find that padding introduces minimal overhead and is required in some cases to maintain the memory alignment imposed by the \mpu APIs.







\subsubsection*{Segment Size $16$:}

\begin{figure}[t]
  \centering
  \includegraphics[width=0.4\textwidth]{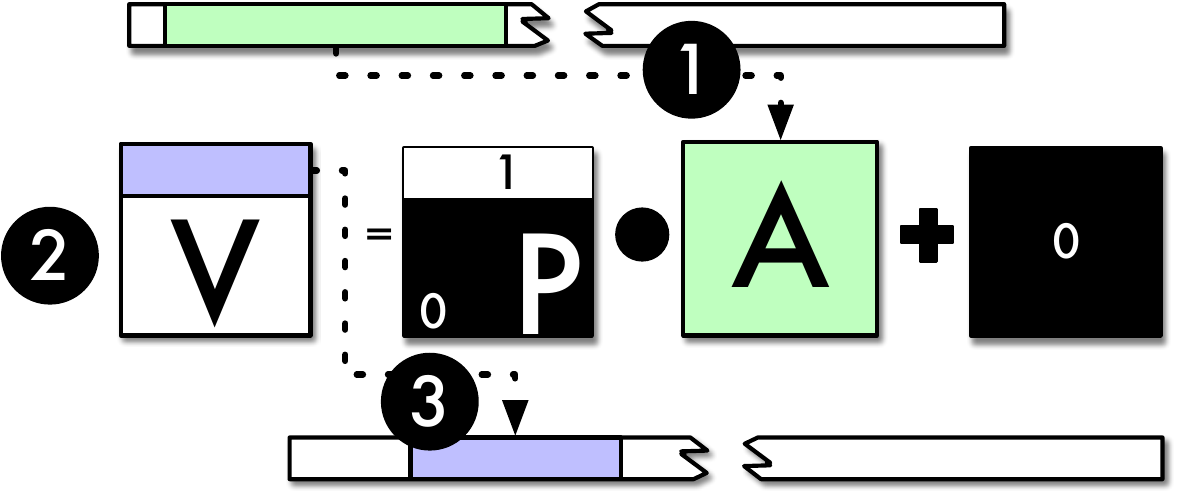}
  \caption{The $Reduction_{16}$ algorithm \circled{1} each warp loads $256$ elements into the matrix $A$ in column major order from the input vector, \circled{2} performs the \mpu operation where the $P$ matrix has ones for the first row, and then \circled{3} the result, which is in the first row of $V$, is stored into the output vector.}
  \label{fig:seg_red_16}
\end{figure}

The $Reduction_{16}$ algorithm, shown in Algorithm~\ref{alg:seg_red_16} and Figure~\ref{fig:seg_red_16}, performs warp-level reduction on $256$ elements which represent $16$ segments of size $16$.
On Line~\ref{alg:seg_red_16:1} in Algorithm~\ref{alg:seg_red_16}  or Step~\circled{1} in Figure~\ref{fig:seg_red_16}, the data is loaded from memory into a column-major order fragment (matrix $A$).
Each row is then reduced using $V = P \cdot A$ (Line~\ref{alg:seg_red_16:2} or Step~\circled{2}).
The result --- first row of $V$ --- is stored in the output memory (Line~\ref{alg:seg_red_16:3} or Step~\circled{3}).

% ... 

\begin{algorithm}
    \caption{The $Reduction_{16}$ algorithm.}
    \label{alg:seg_red_16}
    \begin{algorithmic}[1]
        \State Initialize $P$ matrix.
        \State $idx \gets \textbf{global offset}$
        \State $A \gets \textbf{LoadTile}\left (in \left [idx \ldots idx + 256 \right], ``colmajor" \right )$ \label{alg:seg_red_16:1}
        \State $V \gets P \cdot A + \underline{\mathbf{0}}$ \label{alg:seg_red_16:2}
        \IIf{$laneIdx < 16$}  $out\left [\frac{idx}{16}  + laneIdx \right ] \gets V \left [ laneIdx \right ]$ \label{alg:seg_red_16:3}
    \end{algorithmic}
\end{algorithm}



\subsubsection*{Segment Size 256:}

\begin{figure}[hb]
    \centering
    \includegraphics[width=0.45\textwidth]{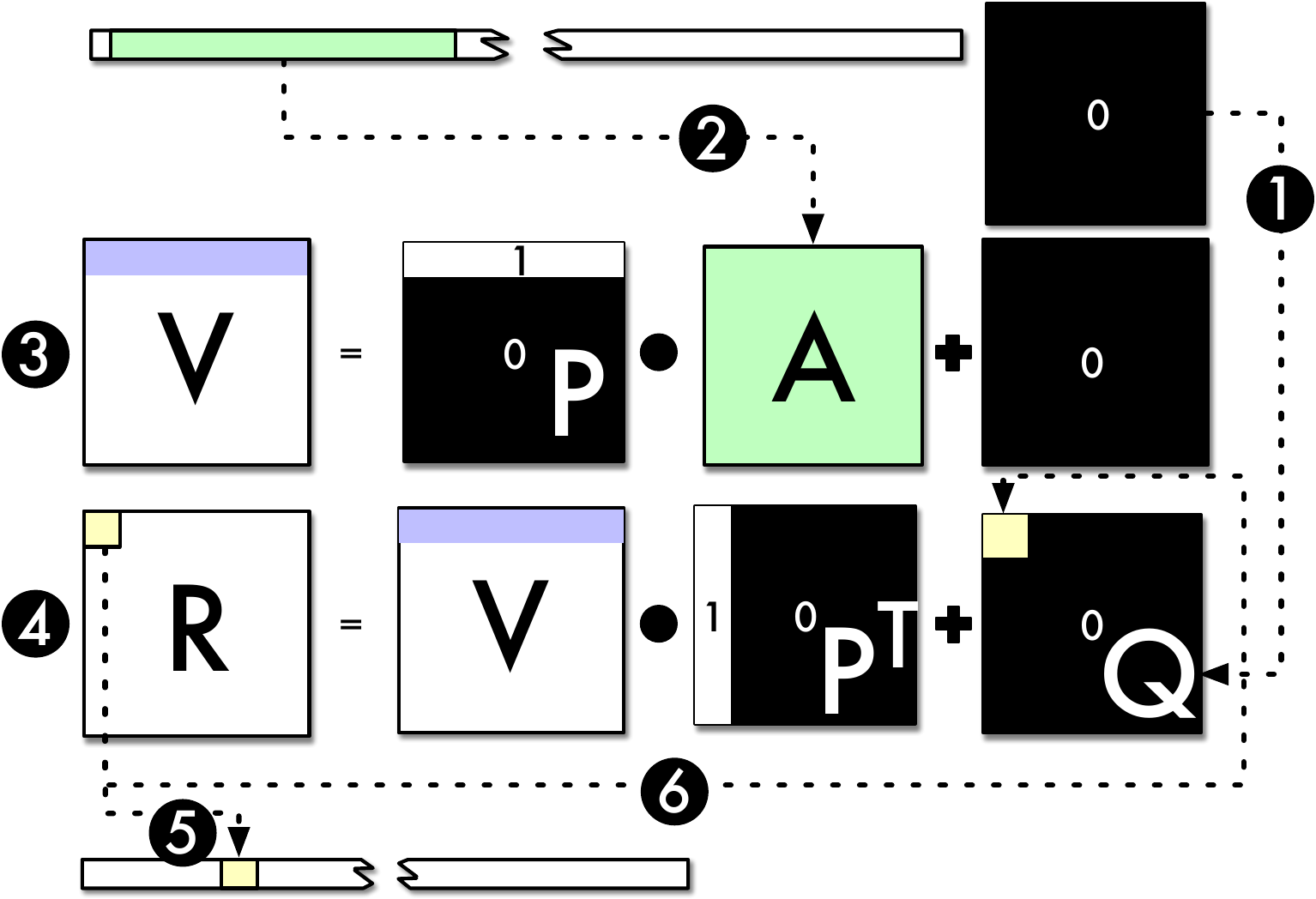}
    \caption{
        The work-inefficient $Reduction_{256N}$ algorithm \circled{1} initializes the $Q$ matrix with all zeros and \circled{2} loads the $256$ input elements into a matrix $A$ in column major order. \circled{3} A dot product $V = P \cdot A + \underline{\mathbf{0}}$ where the $P$ matrix has the first row as ones and the rest of the values are zeros is performed to reduce each row into a scalar. \circled{4} the  dot product $R = V \cdot P^T + Q$ reduces the first row into a scalar. \circled{5} If the segmented reduction size is equal to the matrix size (i.e. $N = 1$) or for the last iteration, then the first element of the $R$ matrix is stored in the output array, otherwise \circled{6} the first element of $R$ is used as the first element of the $Q$ matrix and the procedure is iterated starting from step \circled{2}.
    }
    \label{fig:seg_red_256}
\end{figure}

For handling segments of size $256$, one follows a pattern similar to  $Reduction_{16}$.
The algorithm is shown in Algorithm~\ref{alg:seg_red_256} and is a single iteration of the algorithm illustrated in Figure~\ref{fig:seg_red_256}.
First, all $256$ elements are loaded onto the \mpu (Line~\ref{alg:seg_red_256:1}).
The rows are reduced using the same procedure as $Reduction_{16}$ (Line~\ref{alg:seg_red_256:0}-\ref{alg:seg_red_256:2}) the resulting columns are reduced using $P^T$ (Line~\ref{alg:seg_red_256:3}) before we store the scalar result (Line~\ref{alg:seg_red_256:4}) into memory.

\begin{algorithm}
    \caption{The $Reduction_{256}$ algorithm.}
    \label{alg:seg_red_256}
    \begin{algorithmic}[1]
        \State Initialize $P$ matrix
        \State $idx \gets \textbf{global offset}$\label{alg:seg_red_256:0}
        \State $A \gets \textbf{LoadTile}\left (in \left [idx \ldots idx + 256 \right ], ``colmajor" \right )$ \label{alg:seg_red_256:1}
        \State $V \gets P \cdot A + \underline{\mathbf{0}}$ \label{alg:seg_red_256:2}
        \State $V \gets V \cdot P^T + \underline{\mathbf{0}}$ \label{alg:seg_red_256:3}
        \IIf{$laneIdx = 0$} $out \left [\frac{idx}{256} \right ] \gets V \left [0 \right]$ \label{alg:seg_red_256:4}
    \end{algorithmic}
\end{algorithm}

% ... 
\subsubsection*{Segment Size Multiples of 256:}

With the above $Reduction_{16}$ and $Reduction_{256}$ warp-level primitives, we can express segments that are multiples of either $16$  (denoted by  $16N$) or $256$ (denoted by $256N$).
We will first look at the $256N$ algorithm, since it will be used for representing the $16N$ algorithm.

\begin{figure}[th]
  \centering
  \includegraphics[width=0.45\textwidth]{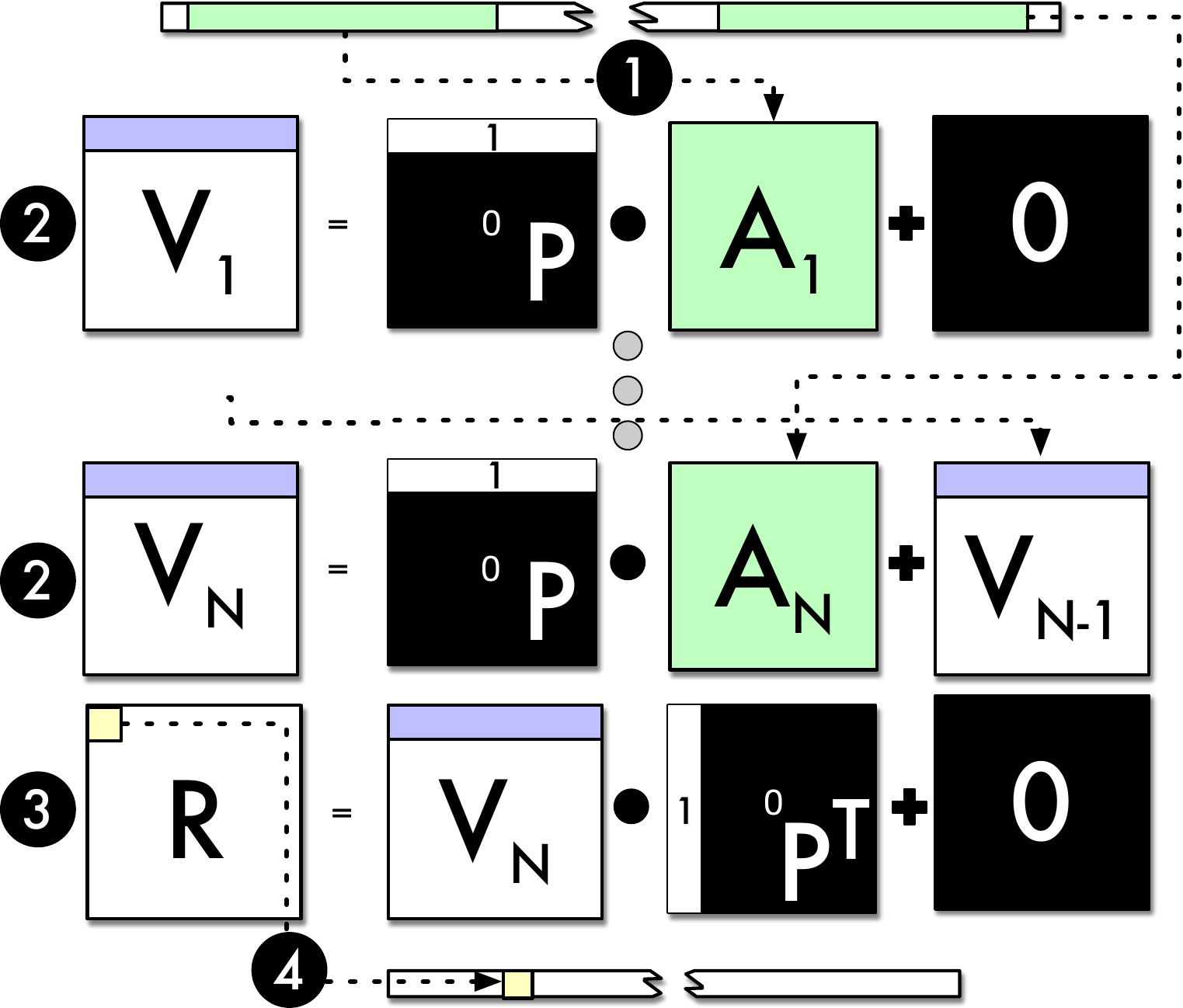}
  \caption{The work-efficient $Reduction_{256N}$ algorithm \circled{1} loads $256$ input elements into matrix $A_i$ in each iteration. It then \circled{2} performs a matrix multiplication $V_i = P \cdot A_i + V_{i-1}$ for $i$ between $1$ and $N$ with $V_0 = \underline{\mathbf{0}}$. The final vector is reduced \circled{3} by performing the $R = V_N \cdot P^T$ operation and the \circled{4} result stored as output.}
  \label{fig:seg_red_256n}
\end{figure}

A na\"ive way is to implement the  $256N$ segmented reduction as $N$-repeated applications of the $Reduction_{256}$, shown in Figure~\ref{fig:seg_red_256}.
While this is correct, it is work inefficient --- wasting one matrix multiplication for each iteration.
Instead of performing two reductions in each iteration, we can implement a work efficient  $256N$ segmented reduction by first reducing each row of the $16 \times 16$ matrix ($Reduction_{16}$) in each iteration and then using the row of reduced values as an accumulator.
In the final iteration, the final row is reduced into a scalar.
Figure~\ref{fig:seg_red_256n} illustrates the work-efficient algorithm.

\subsubsection*{Segment Size Multiples of 16:}

\begin{figure}[t]
  \centering
  \includegraphics[width=0.45\textwidth]{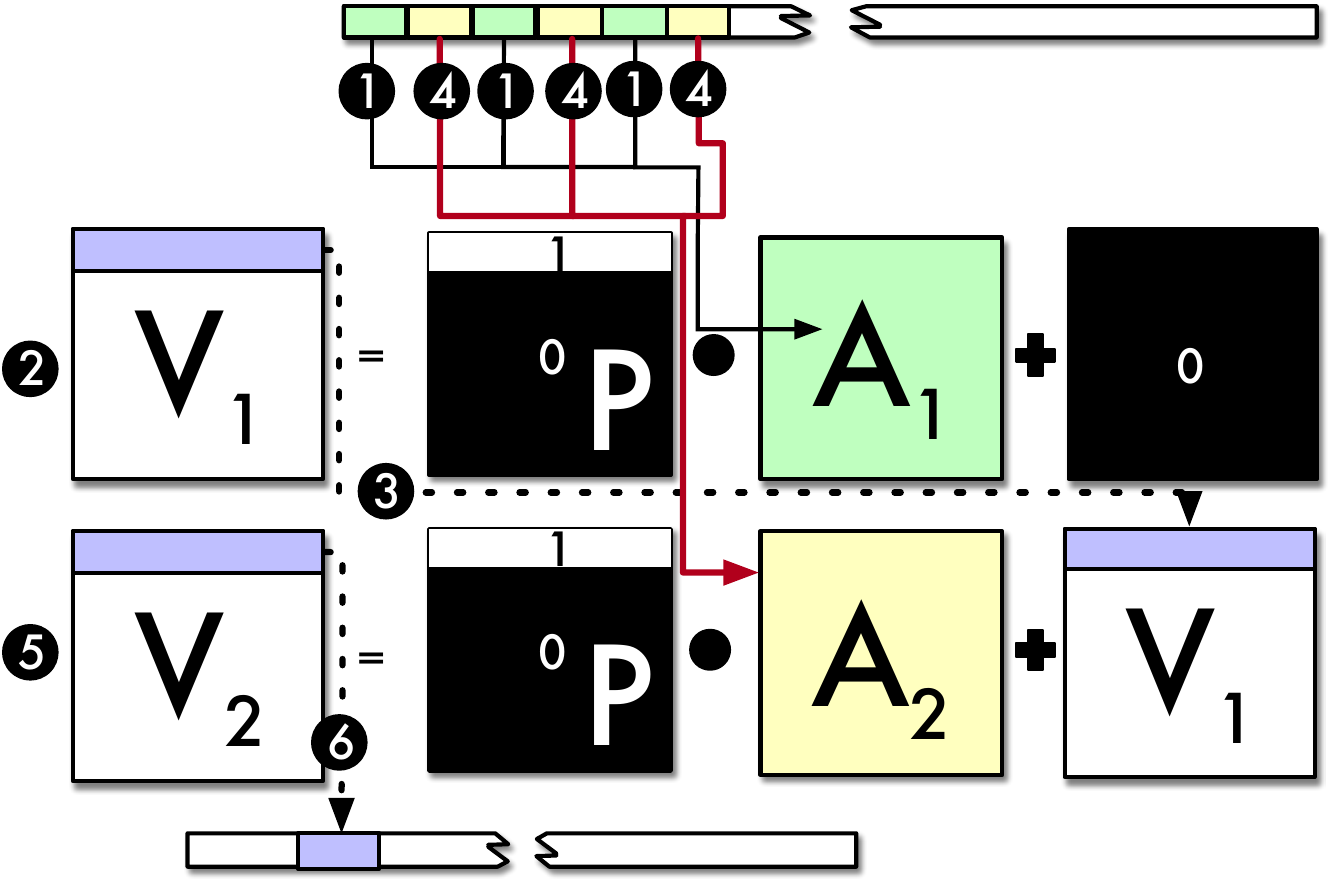}
  \caption{A strided $Reduction_{16N}$ algorithm for $N = 2$  \circled{1} loads $256$ elements where the stride between each row is $16N$. \circled{2} We then perform the matrix multiplication $V_1 = P \cdot A_1$ and \circled{3} use the $V_1$ matrix as an accumulator for the next iteration where \circled{4} we again load the next $256$ elements with the leading dimension set to $16N$. The \circled{5} matrix multiplication $V_2 = P \cdot A_2 + V_1$ is performed and \circled{6} the first row is stored in the output vector.}
  \label{fig:seg_red_16n_2}
\end{figure}

Similar to $Reduction_{256}$, segmented reduction where the segment size is multiples of $16$ ($16N$) can be performed in two ways.
The first is a strided segmented reduction, shown in Figure~\ref{fig:seg_red_16n_2} (for the case where $N = 2$).
During each iteration $i$, a warp loads $16$ segments (each of length $16$) into the matrix $A$ with a stride of $16N$ (Steps~\circled{1} and \circled{4}), i.e., the beginning of each 16-element segment is 16N elements away from the beginning of the next segment in the original input vector.
The $16$ columns of $A$ are then reduced and accumulated into the first row of $V$(Steps~\circled{2} and \circled{5}). 
This repeats for $N$ iterations.
This method is simple, works for arbitrary multiple of $16$, and leverages GPU cache for small $N$.
For large $N$ this method suffers from bad locality.

Algorithm~\ref{alg:seg_red_coalesced_16n} makes better use of cache locality and reduces uncoalesced memory accesses.
The algorithm implements $Reduction_{16N}$ in terms of $Reduction_{256N}$ for $N > 256$.
The left over, $Reduction_{\left (N \% 16 \right ) \times 16}$, can be implemented using the strided segmented $16N$ reduction method.


\begin{algorithm}
    \caption{The Coalesced $Reduction_{16N}$ algorithm.}
    \label{alg:seg_red_coalesced_16n}
    \begin{algorithmic}[1]
        \State Initialize $P$ matrix
        \State $V \gets \underline{\mathbf{0}}$
        \State $gidx \gets \textbf{global offset}$
        \State $numSegs \gets \floor{\frac{16N}{256}}$ \Comment{Number of 256 segments}
        
        \For{$i \gets 0 ; i < numSegs ; i \gets i + 1$}
            \State $idx \gets gidx + 256i$
            \State $A \gets \textbf{LoadTile}\left (in \left [idx \ldots idx + 256 \right ], ``colmajor" \right )$
            \State $V \gets P \cdot A + V$
        \EndFor
        \State $\ldots$ \Comment{Reduce rest of segments using $StridedReduction_{16}$}
        \IIf{$laneIdx < 16$} $out\left [\frac{gidx}{16N} + laneIdx \right ] \gets V \left [laneIdx \right ]$ 
    \end{algorithmic}
\end{algorithm}

\begin{algorithm}
    \caption{The Block-level $Reduction_{256N}$ algorithm.}
    \label{alg:block_red}
    \begin{algorithmic}[1]
        \State $wpb \gets \textbf{warps per block}$
        \State $prtls \gets \textbf{alloc shared mem}\left [wpb \right ]$
        \State $partial \gets Reduction_{256\frac{N}{wpb}}\left (in \right )$
        \IIf{$laneIdx = 0$}  $prtls \left [warpIdx \right ] \gets partial$ 
        \State $\textbf{sync threads}$
        \IIf{$warpIdx=0$} $out \left [blockIdx \right ] \gets Reduction_{wpb}\left (prtls \right )$
    \end{algorithmic}
\end{algorithm}

\subsection{PE/Core (Block)-level Reduction}

When the segment size is large, collaborative reduction within a block becomes profitable.
We follow standard practice~\cite{cub} to implement block-level reduction, but differ in that we still use the \mpu to perform reduction on the partially reduced values within a block.
Algorithm~\ref{alg:block_red} shows how warp-level reduction is used to implement the block-level $Reduction_{256N}$ kernel.

\subsection{Device (Grid)-level Reduction}

When the segment size is very large a grid-level reduction might be needed.
A na\"ive grid-level reduction for a list of length $N$ involves two kernel launches. 
The first kernel launch performs a segmented reduction with the output stored in a list of partials.
A second kernel then reduces the partials into a scalar.
Although this algorithm is na\"ive, its performance is on par with the fastest algorithm.

%% file: sec/5-prefixsum.tex
\section{\mpu Scan Algorithm}\label{sec:prefixsum}


It might be less intuitive to represent scan as matrix multiplication.
For a vector $V$ of $256$ elements, we can store it in row-major order within a $16 \times 16$ matrix $A$ --- with $a_{i,j} = V \left \lbrack 16 \left( j-1 \right) + i \right \rbrack$.

\scalebox{.65}{\parbox{.5\linewidth}{%
\begin{align*}
\begin{aligned}
A = \begin{pmatrix}
a_{1,1} & a_{1,2} & \ldots & a_{1,16} \\
a_{2,1} & a_{2,2} & \ldots & a_{2,16} \\
\vdots & \vdots & \ddots & \vdots \\
a_{16,1} & a_{16,2} & \ldots & a_{16,16}
\end{pmatrix}
\end{aligned}
\end{align*}
}}

We notice that a row-wise scan can be obtained by multiplying the matrix $A$ with an upper diagonal matrix --- with the values of the upper diagonals  being $1$ and the rest $0$.

\scalebox{.65}{\parbox{.5\linewidth}{%
\begin{align*}
\begin{aligned}
Row&Scan \begin{pmatrix}
a_{1,1} & a_{1,2} & \ldots & a_{1,16} \\
a_{2,1} & a_{2,2} & \ldots & a_{2,16} \\
\vdots & \vdots & \ddots & \vdots \\
a_{16,1} & a_{16,2} & \ldots & a_{16,16}
\end{pmatrix} = A \cdot U = 
A \cdot
\begin{pmatrix}
1 & 1 & \ldots & 1 \\
0 & 1 & \ldots &1 \\
\vdots & \vdots & \ddots & \vdots \\
0 & 0 & \ldots & 1
\end{pmatrix}=\begin{pmatrix}
a_{1,1} 
& \ldots & \Sigma\sum_{i=1}^{16} a_{1,i} \\
a_{2,1} 
& \ldots & \Sigma\sum_{i=1}^{16} a_{2,i} \\
\vdots 
& \ddots & \vdots \\
a_{16,1} 
& \ldots & \Sigma\sum_{i=1}^{16} a_{16,i} \\
\end{pmatrix}
\end{aligned}
\end{align*}
}}

Similarly, to get the scan of each column one can use a lower diagonal matrix.
We use a strict lower diagonal, i.e. the diagonal is $0$, to get an exclusive scan of each column.

\begingroup
\setlength\abovedisplayskip{0pt}
\scalebox{.65}{\parbox{.5\linewidth}{%
\begin{align*}
\begin{aligned}
ExclusiveColumnScan &\begin{pmatrix}
a_{1,1} & a_{1,2} & \ldots & a_{1,16} \\
\vdots & \vdots & \ddots & \vdots \\
a_{16,1} & a_{16,2} & \ldots & a_{16,16}
\end{pmatrix} = L \cdot A = \\
\begin{pmatrix}
0 & 0 & \ldots & 0 \\
1 & 0 & \ldots & 0 \\
\vdots & \vdots & \ddots & \vdots \\
1 & 1 & \ldots & 0
\end{pmatrix} \cdot
&
A =
\begin{pmatrix}
0 & 0  & \ldots & 0 \\
a_{1,1} & a_{1,2} & \ldots & a_{1,16} \\
a_{1,1} + a_{2,1} & a_{1,2} + a_{2,2} & \ldots & a_{1,16} + a_{2,16} \\
\vdots & \vdots & \ddots & \vdots \\
\displaystyle\Sigma\sum_{j =1}^{15}a_{j,1} & \displaystyle\Sigma\sum_{j =1}^{15}a_{j,2} & \ldots & \displaystyle\Sigma\sum_{j =1}^{15}a_{j,16} \\
\end{pmatrix}
\end{aligned}
\end{align*}
}}
\endgroup

We then use the $L \cdot A$ matrix to create a $G$ matrix where each element $G_{j,i}$ is the reduction of the $j^{th}$ row of $L \cdot A$.
That is, all elements in the $j^{th}$ row of $G$ are of the same value --- the sum of all elements  preceding the $j^{th}$ row of $A$, i.e.
$G_{j,i} = \displaystyle\Sigma\sum_{k=1}^{j-1} \displaystyle\Sigma\sum_{i=1}^{16} A_{k,i}$.
The $G$ matrix can be generated by multiplying $L \cdot A$ with a matrix with all element values set to 1. 
We then add $G$ to the $A \cdot U$ matrix to generate the scan of $V$  --- which is read in linear row-major order. 

\scalebox{.65}{\parbox{.5\linewidth}{%
\begin{align*}
\begin{aligned}
&Scan \left (V \right ) = L \cdot A \cdot \begin{pmatrix}
1 & 1 & \ldots & 1 \\
1 & 1 & \ldots & 1 \\
\vdots & \vdots & \ddots & \vdots \\
1 & 1 & \ldots & 1
\end{pmatrix} + A \cdot U = G + A \cdot U = \\
&\begin{pmatrix}
a_{1,1} & a_{1,1} + a_{1,2}  & \ldots & \displaystyle\Sigma\sum_{i =1}^{16} a_{1,i} \\
a_{2,1} + \displaystyle\Sigma\sum_{i =1}^{16} a_{1,i}  & a_{2,1} + a_{2,2}  + \displaystyle\Sigma\sum_{i =1}^{16} a_{1,i}  & \ldots & \displaystyle\Sigma\sum_{j =1}^{2} \displaystyle\Sigma\sum_{i =1}^{16} a_{j,i} \\
\vdots & \vdots & \ddots & \vdots \\
a_{16,1} + \displaystyle\Sigma\sum_{j =1}^{15} \displaystyle\Sigma\sum_{i =1}^{16} a_{j,i} &
a_{16,1} + a_{16,2} + \displaystyle\Sigma\sum_{j =1}^{15} \displaystyle\Sigma\sum_{i =1}^{16} a_{j,i}
& \ldots & 
\displaystyle\Sigma\sum_{j =1}^{16} \displaystyle\Sigma\sum_{i =1}^{16} a_{j,i}
\\
\end{pmatrix}
\end{aligned}
\end{align*}
}}

Throughout this section we will use $U$ to represent the upper diagonal matrix where the upper diagonal values are one, and use $L$ to represent the strict lower diagonal matrix where the values below the lower diagonal are one --- i.e. $(U)_{r,c} = \begin{cases} 1 \quad \text{if } r >= c \\ 0 \quad \text{if } r < c \end{cases}$ and $(L)_{r,c} = \begin{cases} 1 \quad \text{if } r < c \\ 0 \quad \text{if } r >= c \end{cases}$.

\subsection{L-Cache (Warp)-level Scan}

With the above derivation, we follow a similar structure to Section~\ref{sec:reduction}: first introducing warp-level primitives before presenting the block- and grid-level primitives.
We write $Scan_K$ to represent a $K$ regular segmented scan. 
Since the process of building warp-level, block-level, and grid-level scans from $Scan_K$ is very similar to that of reduction, we will only highlight the key differences.

\subsubsection*{Segment Size 16:}

Is the $RowScan$ equation above and is illustrated in Figure~\ref{fig:seg_prefixsum} as steps \circled{1}, \circled{2}, and \circled{3}.

\subsubsection*{Segment Size 256:}

\begin{figure}[t]
  \centering
  \includegraphics[width=0.45\textwidth]{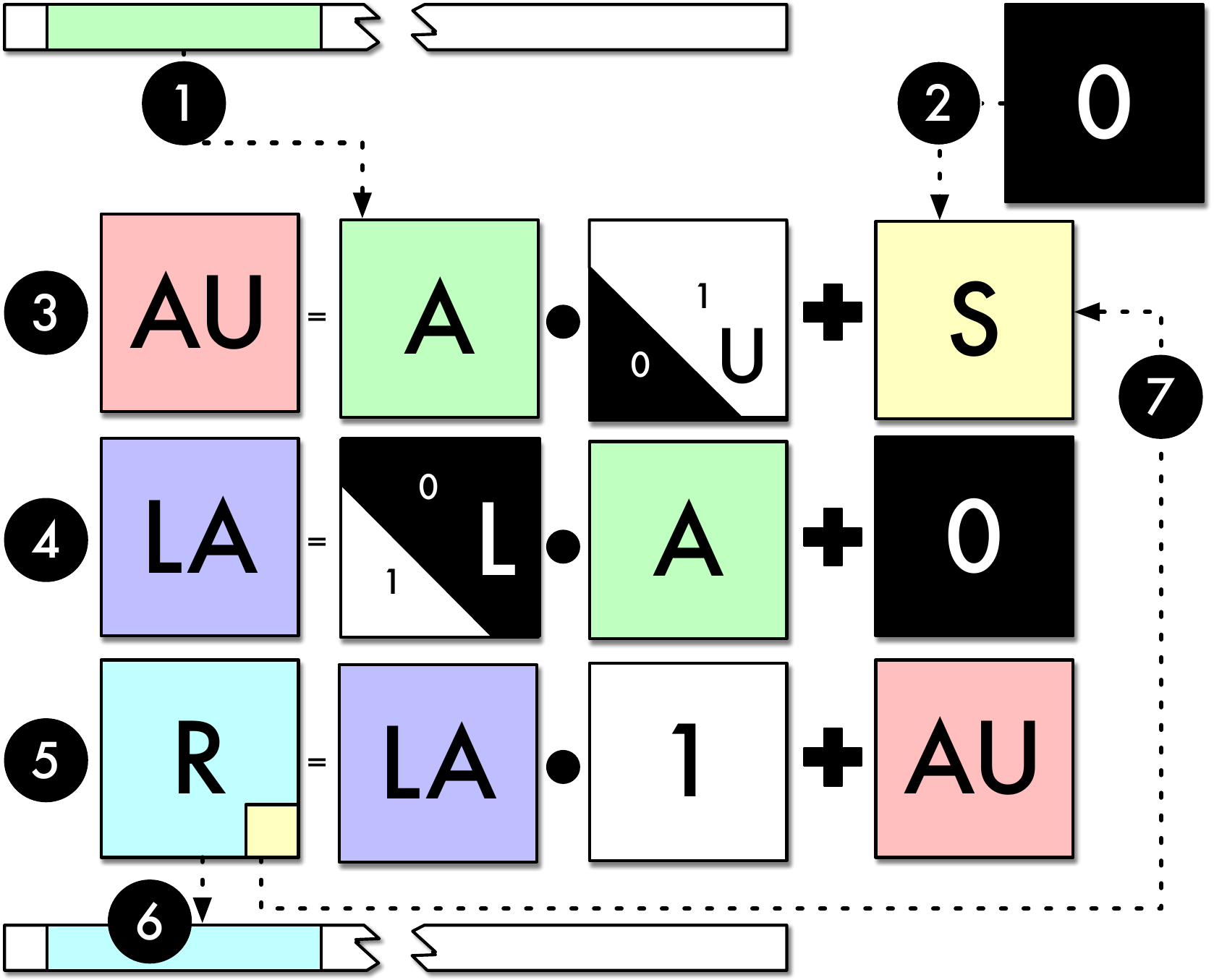}
  \caption{The $Scan_{256N}$ algorithm \circled{1} loads 256 elements from the input vector into a matrix $A$ and \circled{2} initializes the $S$ matrix to $\underline{\mathbf{0}}$. The \circled{3} $AU = A \cdot U + S$ and \circled{4} $LA = L \cdot A + \underline{\mathbf{0}}$ matrix multiplications are performed to compute the prefix sum of each row and column. \circled{5} A row wise reduction is performed on the $LA$ and added to the $AU$ matrix. \circled{6} The result $R$ is stored in the output vector. \circled{7} If the segment size is a multiple of $256$, then the last element of $R$ (position $[16,16]$) is broadcasted into the $S$ matrix and the procedure is repeated.}
  \label{fig:seg_prefixsum}
\end{figure}

Is implemented using $3$ matrix multiplications shown in Figure~\ref{fig:seg_prefixsum} and presented mathematically above.

\subsubsection*{Segment Size Multiples of 16:}

Is similar to strided $16N$ reduction, with the key difference being that we broadcast the last column rather than the reduced scalar value and is shown in Algorithm~\ref{alg:prefixsum_16n}.

\begin{algorithm}
    \caption{The $Scan_{16N}$ algorithm.}
    \label{alg:prefixsum_16n}
    \begin{algorithmic}[1]
        \State Initialize $U$ matrix.
        \State $gidx \gets \textbf{global offset}$
        \State $S \gets \underline{\mathbf{0}}$
        \State $lid \gets laneIdx$
        \For{$i \gets 0; i < N ; i \gets i + 1$}
            \State $idx \gets gidx + 16i$
            \State $A \gets \textbf{LoadTile}\left (in \left [idx \ldots idx + 256N \right ], stride=16N \right )$
            \State $R \gets A \cdot U + S$
            \State $S \gets \textbf{Broadcast}\left (\textbf{LastColumn}\left (R \right ) \right )$
            \If{$lid < 16$}  
                \State $oi \gets idx + lid*16N$
                \State $out \left [oi \ldots oi + 16 \right ] \gets R \left [16 lid \ldots 16 lid + 16 \right ]$
            \EndIf
        \EndFor
    \end{algorithmic}
\end{algorithm}

\subsubsection*{Segment Size Multiples of 256:}

Only a small modification to $Scan_{256}$ is needed to implement $Scan_{256N}$ and is illustrated in Figure~\ref{fig:seg_prefixsum} and Algorithm~\ref{alg:prefixsum_256n}.
Line~\ref{alg:prefixsum_256n:11} in Algorithm~\ref{alg:prefixsum_256n} shows that we keep track of the sum (last element of the $R$ matrix) and broadcast it to the $S$ matrix after each iteration.
The $S$ matrix is then used when performing subsequent iterations.

\begin{algorithm}
    \caption{The $Scan_{256N}$ algorithm.}
    \label{alg:prefixsum_256n}
    \begin{algorithmic}[1]
        \State Initialize $U$ and $L$ matrices.
        \State $gidx \gets \textbf{global offset}$
        \State $S \gets \underline{\mathbf{0}}$ \label{alg:prefixsum_256n:1}
        \For{$i \gets 0; i < N ; i \gets i + 1$}
            \State $idx \gets gidx + 256i$
            \State $A \gets \textbf{LoadTile} \left (in \left [ idx \ldots idx + 256 \right ]  \right )$
            \State $AU \gets A \cdot U + S$
            \State $LA \gets L \cdot A + \underline{\mathbf{0}}$
            \State $R  \gets LA \cdot \underline{\mathbf{1}} + AU$
            \State $out \left [idx \ldots idx + 256 \right ] \gets R$
            \State $S  \gets \textbf{Broadcast}\left (R \left [ 255 \right ] \right )$ \label{alg:prefixsum_256n:11}
        \EndFor
    \end{algorithmic}
\end{algorithm}

\begin{algorithm}
    \caption{The Block-level $Scan_{256N}$ algorithm.}
    \label{alg:block_prefixsum_256n}
    \begin{algorithmic}[1]
        \State Initialize $U$ and $L$ matrices.
        \State $gidx \gets \textbf{global offset}$
        \State $wpb \gets \textbf{warps per block}$ \Comment{Assumed to be less than 16}
        \State $sout \gets \textbf{alloc shared mem}\left [256 \times 16 \right ]$  
        \State $prtls \gets \textbf{alloc shared mem}\left [16 \right ]$ \Comment{Partial sums}
        \State $S \gets \underline{\mathbf{0}}$
        \For{$i \gets 0 ; i < N ; i \gets i + warpsPerBlock$}
            \State $idx \gets gidx + 256 \left (i + warpIdx \right )$  \label{alg:block_prefixsum_256n:1}
            \State $A \gets \textbf{LoadTile}\left (in \left \lbrack idx \ldots idx + 256 \right \rbrack \right )$
            \State $AU \gets A \cdot U + S$
            \State $LA \gets L \cdot A + \underline{\mathbf{0}}$
            \State $R  \gets LA \cdot \underline{\mathbf{1}} + AU$
            \State $sout \left \lbrack 256 warpIdx \ldots 256 warpIdx + 256 \right \rbrack \gets R$ \label{alg:block_prefixsum_256n:2}
            \State $\textbf{sync threads}$
            \If{$warpIdx = 0$}
                \State $E \gets \textbf{LoadTile} \left (sout \left [240 \ldots 4096\right ], stride=256 \right )$  \label{block_prefixsum_256n:line:stride} \label{alg:block_prefixsum_256n:3}
                \State $prtls \gets LastColumnScan_{16} \left (E \right )$ \label{block_prefixsum_256n:line:scan16} \label{alg:block_prefixsum_256n:4} \Comment{Exclusive scan}
            \EndIf
            \State $\textbf{sync threads}$
            \For{$j \gets 1 ; j \leq 256 ; j \gets j + warpSize$} \label{alg:block_prefixsum_256n:5}
                \State $it \gets j + laneIdx$
                \State $val \gets sout \left \lbrack 256warpIdx+it \right \rbrack +prtls \left \lbrack warpIdx\right \rbrack$
                \State $out \left [idx + it \right ]  \gets val$
            \EndFor
            \State $S \gets \textbf{Broadcast}\left( prtls \left \lbrack 15 \right \rbrack \right )$ \label{alg:block_prefixsum_256n:6}
        \EndFor
    \end{algorithmic}
\end{algorithm}

\subsection{PE/Core (Block)-level Scan}

Algorithm~\ref{alg:block_prefixsum_256n} shows how to perform the scan at the block level.
It first computes the segmented scan using the warp primitives (Line~\ref{alg:block_prefixsum_256n:1}-\ref{alg:block_prefixsum_256n:2}), stores the reduced values into a partials list (Line~\ref{alg:block_prefixsum_256n:3}), performs a scan on the partial list (Line~\ref{alg:block_prefixsum_256n:4}), and adds the values to the intermediate results to get the output (Line~\ref{alg:block_prefixsum_256n:5}-\ref{alg:block_prefixsum_256n:6}).

Algorithm~\ref{alg:block_prefixsum_256n} also exercises the \mpu to perform the scan on the partially reduced values across tiles.
On Line~\ref{block_prefixsum_256n:line:stride} we use the offset of the last row ($240$) and $256$ as the leading dimension when loading the tile.
This loads the last row of $R$ across tiles into $E$.
Line~\ref{block_prefixsum_256n:line:scan16} then performs an exclusive scan on the last column of the $E$ and stores the results into the list of partials\footnote{
The implementation of $LastColumnScan_{16}$ is performed by loading the last column values into the first row and performing an \mpu version of the exclusive scan algorithm.
Formulating the intermediate operation this way is needed to adhere to the CUDA WMMA API's byte alignment constraint for loading fragments.}.

\subsection{Device (Grid)-level Scan}\label{sec:grid-level-scan}

Similar to reduction, the segmented scan is used as a building block for the grid-level scan.
The grid-level scan uses a text book implementation, scan-then-propagate strategy, and involves $3$ kernel calls.
The first kernel uses segmented scan and produces partially reduced values for each block.
The second kernel performs a scan on the partially reduced values.
The third kernel then uniformly adds the partially scanned values to their corresponding segments.

%% file: sec/6-evaluation.tex
\section{Evaluation}\label{sec:evaluation}

We implemented the algorithms presented in Sections~\ref{sec:reduction} and \ref{sec:prefixsum} using NVIDIA's WMMA API.
The code (available at \url{https://github.com/c3sr/tcu_scope}) is implemented as a C++ header library with an API similar to CUB's --- providing functions such as \texttt{SegmentedReduce}, \texttt{Reduce}, \texttt{SegmentedScan}, and \texttt{Scan}.
We employ auto-tuning to select the ideal algorithm, number of warps (or independent \mpu operations) per block, coarsening factor (the number segments to perform within a warp), and block dimensions for the user-provided segment size.

We evaluate our implementation on an Intel Xeon E5-2698 with CentOS $4.3$, CUDA Driver $396.26$, and CUDA Runtime $9.2.88$ installed.
We use the Tesla V100-PCIE GPU with 16GB of GPU HBM2 memory and a theoretical peak bandwidth of $900 GB/s$ or $450$ billion half precision elements per second.
All the results below show the throughput of the algorithms in terms of billions of half precision elements per second.


\subsection{Relaxing the WMMA API Constraints}\label{sec:unsafe}

Constraints arise when using the current WMMA API for non-GEMM computation.
These limitations would not exist if one is to perform just GEMM computation. 
The constraints observed were:

\begin{enumerate}
\itemsep0em 
    \item Loads or stores must be performed at fragment granularity.
    \item Loading and storing fragments can only be performed using global or shared memory; constant memory cannot be used.
    \item The memory layout for the matrix kinds are not the same and their is no API to perform casts between them. 
\end{enumerate}

We address these limitations in different ways within our implementation.
For (1) and (2) we use knowledge about the internal layout of the fragment ~\cite{jia2018dissecting} and implemented WMMA API enhancements tailored to our usage.
Listing~\ref{lst:mpu_layout} shows an example of our API extensions for operating on partial fragments.

\begin{lstlisting}[
  float=htp,
  floatplacement=tbp, 
  basicstyle=\fontsize{6}{6}\ttfamily,
  caption={The WMMA API 
  can only perform load/store from shared or global memory and 
  lacks the ability to fill an \mpu fragment from constant memory or operate on sub-fragments. This code shows how we enhance the NVIDIA WMMA API, using knowledge of the fragment layout, to create an upper triangular matrix and get the first column of a fragment for the \texttt{matrix\_b} fragment kind.},
  frame=lines,
  label=lst:mpu_layout,
  captionpos=b]
using frag_b = fragment<matrix_b, 16, 16, 16, half, row_major>;
__device__ int matrix_b_get_row_idx() {
    const int laneIdx = threadIdx.x % warpSize;
    return laneIdx&0x10 >> 2 + laneIdx&0x0B;
}
__device__ void matrix_b_set_upper_triangular(frag_b &f) {
#pragma unroll
    for (int ii = 0; ii < f.num_elements; ii++)
      f.x[ii] = matrix_b_get_row_idx() < ii ? 0.0f : 1.0f; }
__device__ void matrix_b_get_first_column(half* out, frag_b f) {
    const int laneid = threadIdx.x % warpSize;
    if (laneid & 0x04) return ; // avoid redundant writes
    out[matrix_b_get_row_idx()] = f.x[0]; 
}
\end{lstlisting}

Although we can use the layout information to shuffle registers to address (3), we opt instead to express the cast in terms of load/store APIs available through the WMMA API.
For example, to cast a matrix in the \texttt{matrix\_a} format to \texttt{matrix\_b} format, we first store the matrix into shared memory and then perform a load from memory to \texttt{matrix\_b}.
Using our API extensions for fragment layout information requires less block synchronization --- which increases the performance of our implementation by up to $5\%$.
Since relying on fragment layout information is not portable, we omit these results.




\subsection{Optimizing CUB for Half Precision}

CUB is a C++ template library that contains multiple algorithms for the collectives.
The CUB library contains the fastest~\cite{cub2,merry2015performance} implementation for the reduction and scan collectives and is used by libraries such as Thrust~\cite{thrust} as well as most deep learning frameworks~\cite{caffe2,mitchell2017accelerating,mxnet,pytorch,tensorflow}.
We compare against the latest release of CUB~\cite{cub} (version $1.8$) and evaluate against different parameters of the collectives.
As an optimization, warp-level shuffle reduction and scan are implemented in PTX within CUB for integer, float, and double data types, since NVCC is currently unable to use the shuffle instruction's predicate to guard against invalid peers~\cite{shuffle,cudahandbook}.
We observerved that CUB does not contain these shuffle-based optimizations for half precision.
To make the evaluations fair and meaningful, we implement these optimization for the half precision data type in CUB.
The modified CUB is used for the evaluation to provide a more aggressive base of comparison.




\subsection{Warp- and Block-level Reduction and Scan}


Theoretically (on V100) our warp-level \mpu  reduction algorithms require less than one fourth of the cycles of the warp-level shuffle reduction.
For example, consider performing a warp-level $Reduction_{256}$: the warp-level reduction shown in Listing~\ref{lst:warp_collectives} requires $8$ iterations of $32$ element reduction to reduce each segment.
The total cycles is therefore $256$, since each shuffle instruction and addition takes $4$ cycles.
Our algorithm performs the reduction using two matrix multiplications or $64$ cycles --- since each \mpu WMMA matrix multiplication requires  $32$ cycles. 
However, reduction is known to be memory bound, with the ideal performance bounded by memory copy speed.

\begin{figure}[t]
  \centering
   \includegraphics[width=0.45\textwidth]{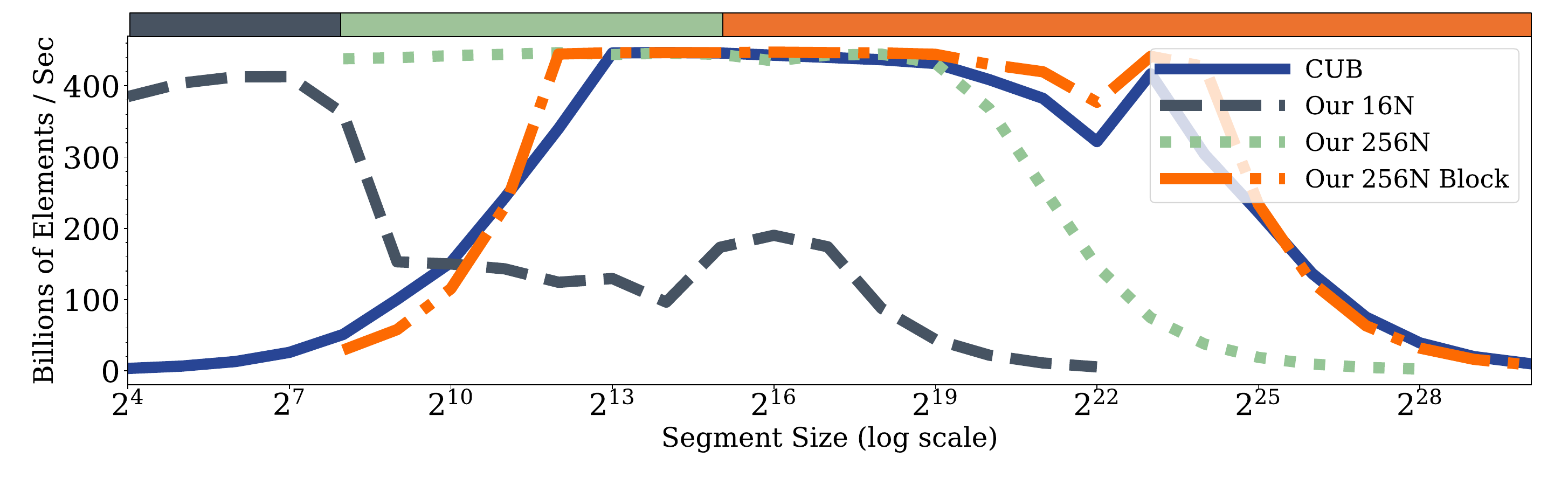}
  \caption{We evaluate the segmented reduction for the algorithms presented on different segment sizes (between $16$ and $2^{30}$) for a fixed $2^{30}$ element list. Through a combination of the algorithms presented, for the range between $16$ and $2^{24}$ we are able to achieve throughput within $90\%$ and $98\%$ of ideal throughput (the theoretical peak is $450$ billion half precision elements per second). The bar on top of the figure shows the best performing algorithm for each range of segment sizes.}
  \label{fig:tune_seg_red}
\end{figure}

We evaluate the \mpu segmented reduction algorithms against \texttt{cub::DeviceSegmentedReduce::Sum} by fixing the number of input elements and varying the segment size (Figure~\ref{fig:tune_seg_red}). 
When the segment size is less than $256$, the $16N$ algorithm is used.
The $16N$ algorithm's performance degrades for large segment sizes due to its strided access pattern resulting in uncoalesced global memory access. 
When the segment size is larger than $256$, the $256N$ algorithm is used, but again suffers from performance degradation after segment size $2^{15}$ due to low occupancy.
When the segment size is large (greater than $2^{15}$) the block-level $256N$ reduction is used. 
Figure~\ref{fig:tune_seg_red} shows that our \mpu implementation achieves more than $90\%$ of the peak throughput  for variable segment size and is always better than CUB.

\begin{figure*}[ht!]   
\centering
\includegraphics[width=0.95\textwidth]{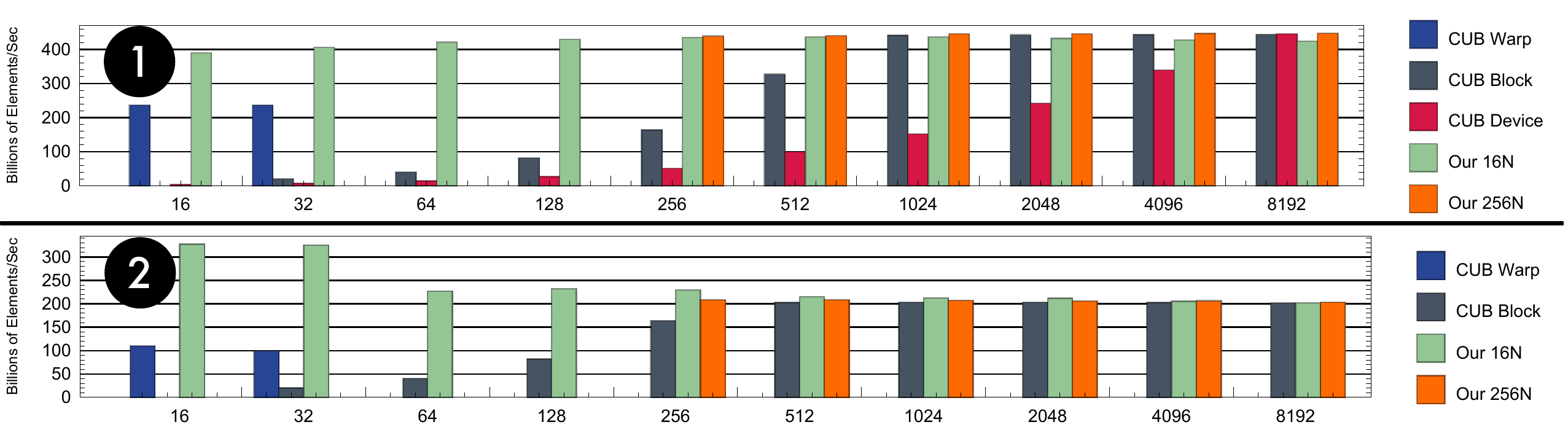}
\caption{
Segmented \circled{1} reduction and \circled{2} scan are evaluated in terms of billions of half-precision elements per second ($y$-axis) for segment sizes between $2^4$ and  $2^{13}$ ($x$-axis). The best  configurations for our implementation as well as CUB are selected.
}
\label{fig:against_cub}
\end{figure*}

When the segment size is large and the number of segments is small, the performance of both CUB and our implementation drops. 
Since each segment gets mapped onto a block, a small number of segments causes some SMs to be idle. 
For example when segment size is $2^{25}$, both CUB and our implementation achieve an occupancy of around $0.25$ and SM efficiency of around $40\%$.
A better strategy for these cases would be to assign multiple thread blocks to collaboratively reduce each segment when the size of the segments is very large.
This optimization can be achieved using CUDA 9's cooperative groups~\cite{coop}, but is outside the focus of this paper.

\begin{figure}[t]
  \centering
  \includegraphics[width=0.45\textwidth]{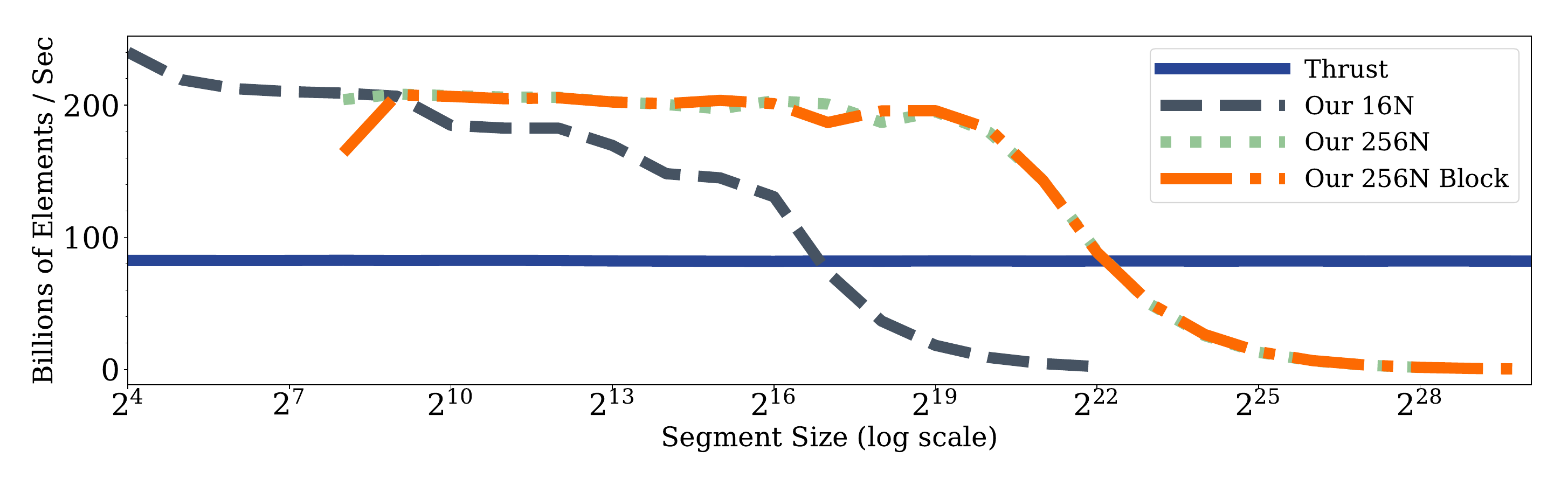}
  \caption{We evaluate the segmented scan for the algorithms presented on different segment sizes for a fixed $2^{31}$ element list. Through a combination of the algorithms presented, for the range between $16$ and $2^{19}$ we are able to achieve throughput within $89\%$ and $97\%$ of ideal throughput (the theoretical peak is $225$ billion half precision elements per second).}
  \label{fig:tune_segmented_scan}
\end{figure}

Our \mpu implementation largely outperforms CUB's device-wide segmented reduction for different segment size.
Through profiling, we identified the major predictors of performance to be, in the order of importance, the number of half-precision floating-point instructions (\texttt{inst\_fp\_16} in the NVProf~\cite{nvprof} metrics), warp instructions (\texttt{inst\_inter\_thread\_communication}), and integer instructions (\texttt{inst\_integer}).
We consistently find that our implementation's half-precision instructions is approximately equal to the number of total elements ($2^{31}$) while CUB's is much larger.
Moreover, CUB requires large number of integer  and warp shuffle  instructions while our implementation uses no warp shuffle instructions and a smaller number of integer instructions.
This contributes to the $100\times$ speedup for segment size $16$.

We examined the power consumption by measuring the average power draw within the execution phase of the kernel using NVProf.
Based on these measurements, we find that our implementation consumes $7.4-22.3\%$ less power compared to CUB across different segment sizes.
Again, this is because of the efficient use of the $FP16$ and $INT$ ALUs as well as better SM and DRAM utilization.
We note that our algorithm leaves the general purpose ALUs idle, allowing less contention on these units.

CUB provides a \texttt{cub::WarpReduce}, applicable for segment sizes $16$ and $32$, to compute a parallel reduction of elements within a warp.
CUB also provides \texttt{cub::BlockReduce} to perform reduction within a block.
These two primitives require users to partition the data and construct the kernel.
Since CUB's device-wide segmented reduction does not perform well for segment size smaller then $2^{13}$, we evaluate our \mpu implementations against \texttt{cub::WarpReduce} and \texttt{cub::BlockReduce} implementations, shown in Figure~\ref{fig:against_cub}.
The \texttt{cub::WarpReduce} implementation is tunable on block size, wheras the \texttt{cub::BlockReduce} implementation is tunable on block size, thread coarsening factor, and reduction algorithms.
We compare our implementation against the best CUB implementation.
We find that our \mpu implementations is still faster for segment size smaller than $1024$, and is comparable to \texttt{cub::BlockReduce} for the other cases.


For segmented scan, we evaluate the \mpu algorithms against Thrust's implementation (\texttt{inclusive\_scan\_by\_key}), since CUB has no user visible API for segmented scan. 
The Thrust implementation utilizes CUB's internal warp- and block-level scan to implement the scan-by-key operation.
We evaluate different segment sizes with a fixed number of input elements --- the results are shown in Figure~\ref{fig:tune_segmented_scan}.
Thrust, consistent with previous work~\cite{eilers2014multireduce}, has constant performance irrespective of the segment size.
Whereas, our scan \mpu implementations achieve more than $89\%$ of the peak throughput and is $3\times$ faster than thrust for small segment sizes.
We observe lower power consumption compared to Thrust --- observing it to be either equivalent in power usage or up to $17\%$ less.
Our segmented scan is not ideal for large segment sizes since, as explained in Section~\ref{sec:reduction},   only a small number of blocks get launched and thus the GPU is underutilized.
This inefficiency can be remedied using the same strategy described for reduction.

\begin{figure}[t]
  \centering
  \includegraphics[width=0.45\textwidth]{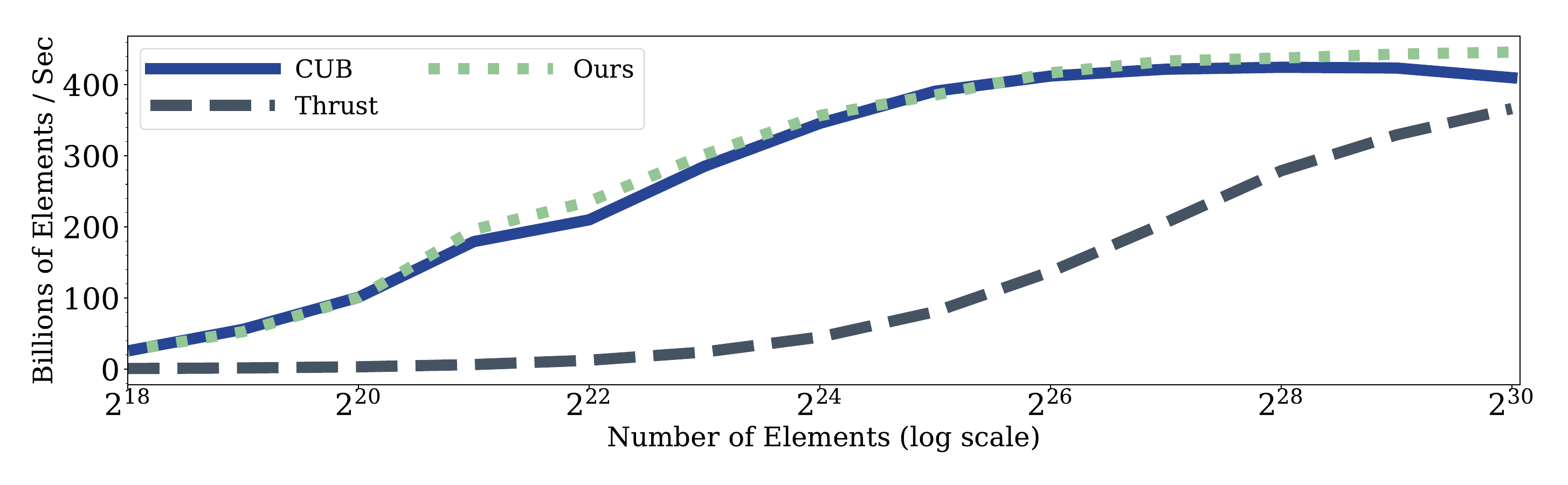}
  \caption{A full reduction implementation based on the description in Section~\ref{sec:reduction} achieves performance on par to CUB.
  }
  \label{fig:eval_full_reduction}
\end{figure}

CUB provides \texttt{cub::WarpScan} to compute a parallel scan of data partitioned within a warp, and \texttt{cub::BlockScan} within a block.
Similar to reduction, these two primitives require more programming effort from the users to partition the data and construct the kernel. 
The CUB scan implementations have the same tunable parameters as CUB's reduction. 
We evaluate our \mpu segmented scan against the best \texttt{cub::WarpScan} and \texttt{cub::BlockScan} parameters, shown in Figure~\ref{fig:against_cub}.
We can see that our \mpu implementations are still faster for small segment size, and are at least comparable to \texttt{cub::BlockScan} for other cases.


\subsection{Grid-level Reduction and Scan}

Unlike the warp- and block-level operations, this paper does not attempt to optimize grid-level operations --- opting to use a na\"ive implementation for the grid-level collectives.
The na\"ive implementation involves multiple kernel launches.
We include the evaluation results to show that even our  na\"ive grid-level implementation achieves performance that is better or comparable to that of CUB and Thrust.

\begin{figure}[h]
  \centering
  \includegraphics[width=0.45\textwidth]{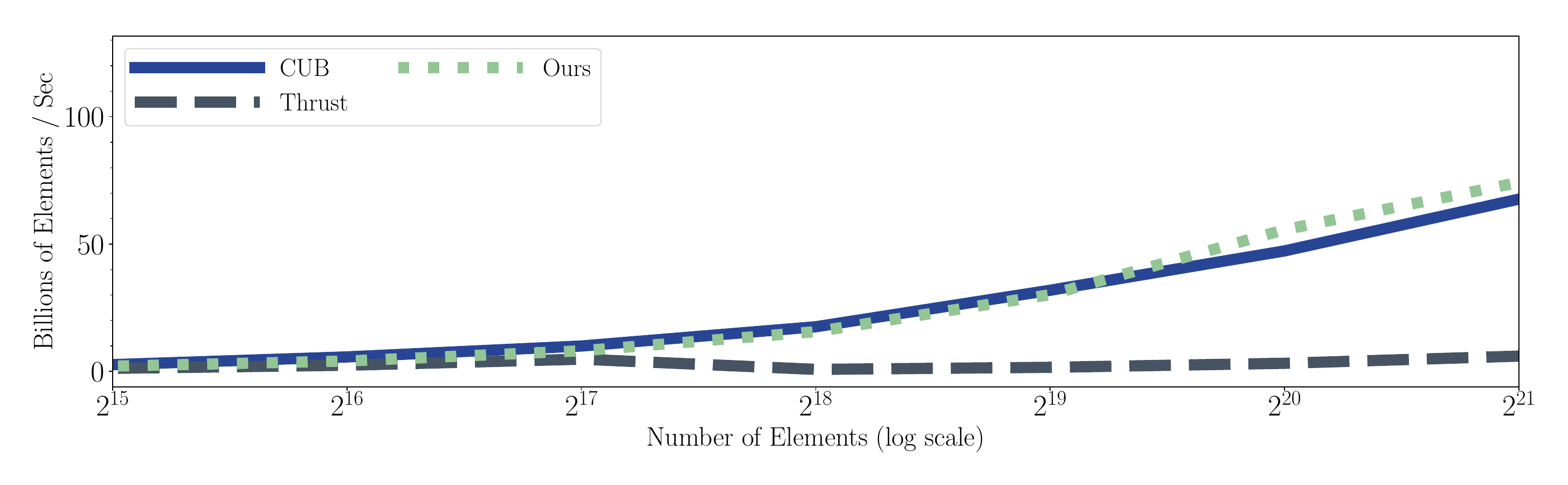}
  \caption{A full scan implementation based on the description in Section~\ref{sec:prefixsum} achieves performance comparable to CUB.
  }
  \label{fig:eval_full_scan}
\end{figure}

We compare against both CUB and Thrust for full reduction (Figure~\ref{fig:eval_full_reduction}), and scan (Figure~\ref{fig:eval_full_scan}).
For both cases, our implementation uses the $256N$ block-level algorithm.
Even with our na\"ive grid-level implementation, we are able to mach  the performance of CUB and are considerably faster than the Thrust implementation. 
For reduction and scan, the TCU implementation is slightly faster than CUB with large input sizes being bounded by memory bandwidth and is within $98\%$ (for reduction) of peek memory copy bandwidth.
For scan, our current unoptimized implementation uses CUB to reduce the partial values (kernel 2 described in Section~\ref{sec:grid-level-scan}). 
Future implementations would not use CUB, since it fails for inputs larger than $2^{21}$ which causes our implementation to fail as well.

%% file: sec/8-related.tex
\section{Related Work}\label{sec:related}

The mapping of algorithms onto matrix multiplication has been well studied~\cite{huang1998fast,kolda2009tensor,rabanser2017introduction,van1992survey}.
Similarly, both reduction and scan are well studied from a performance and application~\cite{blelloch1989scans,blelloch1993segmented,harris2007parallel,kim2011scalable,sengupta2006work} aspect on a wide range of architectures and have been heavily evaluated across GPU architectures~\cite{dotsenko2008fast,dybdal2016low,mcdonell2013optimising,sengupta2008efficient,yan2013streamscan}.
To our knowledge however, there has been no attempt at mapping either reduction or scan in terms of matrix multiplication.

Considerable research has been done on the development of performance portable compilers for  reduction and scan kernels~\cite{petabricks,tangram,de2019automatic,larsen2017strategies,lift}.
These compilers express the algorithms as systems of alternative building blocks that are then composed and auto-tuned at compile time for both the target architecture and the input characteristics.
These tools are synergistic with our technique, since we are able to add our algorithm as another building block to implement reduction or scan.

Previous work~\cite{chaurasia2015compiling,cub,merrill2010revisiting,halide,yan2013streamscan} has also shown that optimizations can be made to either avoid or hide the overhead of multi-kernel launches.
These optimizations would enable our grid-level operations to be competitive for large sizes when compared to state-of-the-art methods.
Other research looked at specific cases of scan, in~\cite{sam} the authors look at performing scan on tuples while minimizing global reads and facilitating latency hiding.

Work describing NVIDIA's Tensor Cores is scarce. 
In~\cite{jia2018dissecting}, the authors use microbenchmarks to discern micro-architectural details of the V100 architecture.
This work was extended in~\cite{Raihan} where authors' study expand on the micro-architectural study and show a proposed NVIDIA \mpu architecture.
In~\cite{markidis2018nvidia, haidar2017investigating} the authors use half precision and \mpus to implement iterative solvers.
They use half precision along with low quality solvers to compute the initial conditions and then switch to both higher precision solvers for subsequent iterations.
The authors also examine the numerical error incurred when using \mpus and half-precision for HPC workloads.


%% file: sec/9-conclusion.tex
\section{Conclusion}\label{sec:conclusion}

This paper leveraged the Tensor Core Units (\mpus) (a specialized accelerator developed to optimize matrix multiplication for deep learning) to implement both reduction and scan.
We showed a novel, simple, and efficient mapping of the reduction and scan primitives onto \mpus.
We believe we are the first to formulate these algorithms to exercise the \mpu.
Unlike existing work which designs ASICs to map reduction and scan onto hardware, we develop an algorithmic solution to map both reduction and scan on existing \mpus. 
An algorithmic solution is relevant when using 
preexisting \mpu designs (as is the case for the NVIDIA \mpu).
We also pointed out directions for future API and architectural changes to relax some of the \mpu constraints such as loading fragments from constant, extracting single row or column, etc. --- resulting in a  simplified implementation.

We implemented the proposed algorithms onto V100 \mpus, achieved up to $100\times$ speedup for reduction and up to $3\times$ for scan, and showed performance that rivals state of the art implementation in the worst cases.
We observed up to $22\%$ less power consumption for reduction and $16\%$ for scan using NVPROF.
As a result of the algorithms, we were able to make use of the otherwise idle \mpus --- enabling better GPU utilization for kernels that exercise the general purpose ALUs.

Future work would leverage the techniques described in this paper to map more algorithms and functions onto \mpus. 
We are specifically interested in transcendental and special functions, since the NVIDIA special function units have been observed to be the bottleneck in HPC applications.
We also want to express neural network layers in terms of \mpus, where some layer implementations and layer fusion opportunities would be enabled by our work: such as the computation of variance in batch norm~\cite{ioffe2015batch,2018arXiv180711205J} or the evaluation of special functions in activation layers.


%% file: sec/99-ack.tex

\begin{acks}
\label{sec:ack}

This work is supported by IBM-ILLINOIS Center for Cognitive Computing Systems Research (C3SR) - a research collaboration as part of the IBM Cognitive Horizon Network.

\end{acks}